\documentclass[aps,prb,showpacs,floatfix,twocolumn,superscriptaddress]{revtex4-2}
\usepackage{placeins}
\usepackage{xcolor}
\usepackage{graphicx}
\usepackage{dcolumn}
\usepackage{float}
\usepackage{relsize}
\usepackage{cancel}
\usepackage{bm}
\usepackage{tikz}
\usepackage{siunitx}
\usepackage{tablefootnote}
\usepackage{physics}
\usetikzlibrary{arrows.meta,positioning}

\usepackage[
  pdfauthor={Wanying Zheng and Tao Chen and Youjin Deng and Wanzhou Zhang},
  pdfstartview=FitH,%
  breaklinks=true,%
  bookmarks=true,%
  colorlinks=true,%
  anchorcolor=black,%
  citecolor=red,
  filecolor=black,%
  menucolor=black,%
  urlcolor=blue,%
  linkcolor=red,%
 ]{hyperref}
\usepackage[english]{babel}
\usepackage{amssymb}
\usepackage{color, soul}
\usepackage{amsmath}
\usepackage{amstext}
\usepackage{latexsym}
\usepackage{mathtools}
\usepackage{booktabs}
\begin{document}

\title{Tensor network Monte Carlo study of the $J_1$-$J_2$ Ising model on the square lattice}

\author{Wanying Zheng}
\affiliation{College of Physics and Optoelectronic Engineering, Taiyuan University of Technology, Shanxi 030024, China}
\author{Tao Chen}%
%\thanks{These authors contributed equally.}
\affiliation{
Hefei National Laboratory for Physical Sciences at the Microscale and Department of Modern Physics, University of Science and Technology of China, Hefei 230026, China}

\author{Youjin Deng}%
\email{yjdeng@ustc.edu.cn}
\affiliation{
Hefei National Laboratory for Physical Sciences at the Microscale and Department of Modern Physics, University of Science and Technology of China, Hefei 230026, China}
\author{Wanzhou Zhang}%
\email{zhangwanzhou@tyut.edu.cn}
\affiliation{College of Physics and Optoelectronic Engineering, Taiyuan University of Technology, Shanxi 030024, China}

\begin{abstract}

The frustrated square-lattice \(J_1\)-\(J_2\) Ising model exhibits rich magnetic phases, yet its phase transitions remain difficult to characterize due to critical slowing down in conventional Monte Carlo and truncation errors inherent to tensor renormalization group methods.
We develop a tensor-network Monte Carlo (TNMC) algorithm for this frustrated model, which supports both open and periodic boundary conditions. For periodic lattices, we use a sweeping block-update strategy that updates spin blocks while fixing surrounding spins.
To boost efficiency, we couple our sampler with histogram reweighting, and test each technique standalone to validate these independent optimizations.
We perform extensive finite-size scaling analyses at \(g=0.55\) and \(g=1.0\), extracting apparent critical temperatures and effective thermal exponents from Binder-cumulant data.
Our method offers an efficient, nearly bias-free path for large-scale simulations of \(J_1\)-\(J_2\)-type frustrated magnets under periodic boundaries and generalizes readily to other frustrated spin lattices.
\end{abstract}

\maketitle
\section{Introduction}

Frustrated spin systems are central to statistical condensed matter physics, where competing exchange couplings destroy simple magnetic order and give rise to rich phases and critical phenomena~\cite{frustrated_book}. The square-lattice $J_1$-$J_2$ Ising model, with nearest-neighbor coupling $J_1$ and diagonal next-nearest-neighbor coupling $J_2$ (frustration strength $g=J_2/|J_1|$), serves as a minimal frustrated benchmark~\cite{Moran-Lopez1993,AguileraGranja1993,Moran-Lopez1994,Lopez-Sandoval1999,dosAnjos2008,Kalz2011,jin2012ashkin,jin2013prb,li2021tensor,sun2026generalized}. The phase diagram and critical behavior of this model were mapped out in the early numerical studies of Binder and Landau~\cite{Binder1980,landau1980} and Oitmaa~\cite{Oitmaa1981}. At zero temperature, $g<0.5$ yields ferromagnetism and $g>0.5$ stripe antiferromagnetism, with paramagnetic disorder at high temperature. The critical frustration $g^*$ and the order of the stripe transition have been debated for decades: early mean-field, cluster-variation and small-scale Monte Carlo works placed a broad weak-first-order window at $g^*\in[0.9,1.14]$~\cite{Moran-Lopez1993,AguileraGranja1993,Moran-Lopez1994,Lopez-Sandoval1999,dosAnjos2008}; extensive Monte Carlo simulations then supported weak first order over the wider range $1/2<g\lesssim1$~\cite{Kalz2011}, until finite-size Markov chain Monte Carlo located the continuous region at $g\gtrsim g^*=0.67$~\cite{jin2012ashkin,jin2013prb} and infinite time-evolving block decimation and higher-order tensor renormalization group calculations pushed the first-order line down to $g\lesssim0.54$~\cite{li2021tensor}. The controversy nevertheless persists: a recent matrix-product-state study argues that the stripe transitions are weak first order for $0.67\lesssim g<\infty$, asymptotically becoming second order as $g\to\infty$, and first order for $0.5<g\lesssim0.67$~\cite{gangat2024}; large-scale Monte Carlo simulations find the transitions for $g\geq0.58$ likely continuous~\cite{lijiang2024ptep}; tensor-network renormalization of a symmetry-enhanced relative of the model reports coexisting first- and second-order regimes, with the Ashkin-Teller criticality attached to the second-order line~\cite{chatelain2024}. A field-mixing Monte Carlo analysis~\cite{kalz2012potts,sun2026generalized} locates the endpoint at $g^*\approx0.697$ as a four-state Potts critical point with logarithmic corrections. Recent transfer-matrix~\cite{hucharbonneau2021}, metastable-state~\cite{abalmasov2023}, and competing-interaction~\cite{leekimkim2024} studies continue this debate. Resolving this discrepancy calls for a simulation method that combines large-lattice capability with freedom from sampling bias.

Numerical simulation remains the dominant approach to resolve the phase diagram and critical properties of frustrated Ising lattices, yet two mainstream families of methods suffer unavoidable drawbacks.
First, conventional Markov chain Monte Carlo (MCMC) methods including local Metropolis-Hastings~\cite{Metropolis1953EquationOS}  and classic Swendsen-Wang/Wolff cluster algorithms~\cite{SwendsenWang1987,Wolff1990} are severely limited near critical points. Local single-spin updates are trapped in metastable energy valleys, producing strong temporal autocorrelation and critical slowing down; cluster update schemes fail on frustrated lattices because competing couplings produce poorly sized clusters that  break sampling ergodicity. 
Second, tensor renormalization group algorithms such as the infinite time-evolving block decimation (iTEBD) and higher-order tensor renormalization group (HOTRG)~\cite{li2021tensor,hotrgj1j2023} compute partition functions via direct tensor contraction without stochastic sampling, delivering high-precision equilibrium observables. However, these methods require growing virtual bond dimensions when approaching the frustration boundary, leading to unaffordable memory and computation cost for large system sizes. Standard matrix product state (MPS) periodic boundary implementations further amplify truncation errors due to the absence of a canonical center~\cite{vidal2007itebd}. These bottlenecks motivate the development of hybrid stochastic-tensor simulation frameworks that combine the flexibility of Monte Carlo and the precision of tensor networks.

Among these hybrid approaches, Tensor Network Monte Carlo (TNMC) is particularly promising, with early tensor-network sampling ideas developed in Ref.~\cite{ferris2015tnmc}. It evaluates Boltzmann conditional probabilities via tensor contractions and performs global block spin updates, nearly eliminating critical slowing down~\cite{cut}. Recent work successfully realized TNMC simulations on random-bond disordered Ising lattices, supporting lattice sizes up to \(1024\times1024\) and enabling precise critical exponent extraction without severe sampling correlation~\cite{chen2025tensor,chen2025batchtnmcefficientsamplingtwodimensional}. The framework has further been generalized to three-dimensional classical lattices~\cite{chen2025tensornetworkmarkovchain}, quantum lattice systems~\cite{chen2026variationalmontecarlovmc}, and tensor network population annealing~\cite{oshima2026tensornetworkpopulationannealing} and a unified cluster-updating frame~\cite{guo2024unifiedframeworkclustermethods}. However, existing TNMC tensor constructions are designed only for single nearest-neighbor coupling and cannot be directly transferred to \(J_1\)-\(J_2\) systems; moreover, periodic-boundary treatments for the frustrated \(J_1\)-\(J_2\) lattice are still absent in TNMC simulations, although approximate tensor-contraction sampling under periodic boundaries has been explored for spin glasses~\cite{gangat2024hyperopt}. 

 In the present paper, we generalize the TNMC algorithm to the two-dimensional geometrically frustrated \(J_1\)-\(J_2\) Ising model defined on a square lattice, considering both open (free) boundary conditions and periodic boundary conditions. Numerical simulations are carried out to characterize the computational performance of the TNMC method and extract the intrinsic critical properties of the system at two typical frustration strengths \(g=0.55\) and \(g=1.0\).
 We further couple the TNMC sampler with histogram reweighting~\cite{Ferrenberg1988PRL,Ferrenberg1989PRL} to obtain thermodynamic observables over a finite temperature window from a single simulation, substantially reducing the computational cost.
 We also explore a GPU-accelerated, batch-parallel variant of the sampler as a preliminary performance check.
 
The rest of this paper is organized as follows. Section \ref{sec:method} elaborates the fundamental principles of tensor network Monte Carlo (TNMC), details its implementation for the \(J_1\)-\(J_2\) spin model under both open and periodic boundary conditions, and defines all thermodynamic observables adopted in our simulations and additionally introduces the histogram reweighting technique. Section \ref{sec:res}
 contains two parts. For algorithmic benchmarks, we analyze sampling acceptance probability, autocorrelation functions and computational wall time across various parameters. For physical properties, we extract critical temperatures and critical exponents at typical frustration ratios \(g=0.55\) and \(g=1.0\). Conclusion and discussion are made in Sec.~\ref{sec:con}.

\section{Method}
\label{sec:method}
\subsection{The basic idea of TNMC}
We adopt the frustrated \(J_1\)-\(J_2\) Ising spin model as our benchmark system, whose Hamiltonian takes the form
\begin{equation}\label{eq:Hamiltonian}
E=J_1\sum_{\langle ij\rangle}\sigma_i\sigma_j+J_2\sum_{\langle\langle ij\rangle\rangle}\sigma_i\sigma_j.
\end{equation}
Each lattice site carries an Ising spin \(\sigma_i=\pm1\). The sums \(\langle ij\rangle\) and \(\langle\langle ij\rangle\rangle\) run over nearest-neighbor and next-nearest-neighbor spin bonds, with \(J_1\) and \(J_2\) denoting their respective coupling strengths.

In Monte Carlo simulations, the equilibrium distribution of a physical system obeys the Boltzmann distribution. The probability of the system residing in a spin configuration \(\{\sigma_i\}\) is
\begin{equation}
\pi(\{\sigma_i\}) = \frac{e^{-E(\{\sigma_i\})/T}}{Z},
\label{eq:boltzman}
\end{equation}
where \(T\) is the temperature, the Boltzmann constant is set to unity, and \(Z\) denotes the partition function.

Within the Metropolis-Hastings algorithm, suppose the current spin configuration is \(\sigma_{\text{a}}\) and a trial configuration is \(\sigma_{\text{b}}\). The acceptance probability for transitioning from \(\sigma_{\text{a}}\) to \(\sigma_{\text{b}}\) reads
\begin{equation}
p(\sigma_{\text{b}}\vert\sigma_{\text{a}}) = \min\left\{ 1, \frac{g(\sigma_{\text{a}}\vert\sigma_{\text{b}})}{g(\sigma_{\text{b}}\vert\sigma_{\text{a}})} \cdot \frac{\pi(\sigma_{\text{b}})}{\pi(\sigma_{\text{a}})} \right\},
\label{eq:pa}
\end{equation}
where \(g(\sigma_{\text{b}}\vert\sigma_{\text{a}})\) is the proposal probability of generating \(\sigma_{\text{b}}\) from \(\sigma_{\text{a}}\), and \(g(\sigma_{\text{a}}\vert\sigma_{\text{b}})\) is the reverse proposal probability. This transition rule strictly satisfies the detailed balance condition.

Following the TNMC formulation of Refs.~\cite{cut,chen2025tensor}, the proposal probability possesses a specialized form that depends \textit{only} on the target configuration \(\sigma_{\text{b}}\), i.e., \(g(\sigma_{\text{b}}\vert\sigma_{\text{a}})\equiv g(\sigma_{\text{b}})\). It can be further decomposed into a product of sequential conditional probabilities:
\begin{equation}
g(\sigma_{\text{b}}) = \prod_{i=1}^N P(\sigma_i\vert\sigma_{<i}).
\end{equation}
Each conditional probability \(P(\sigma_i\vert\sigma_{<i})\) is defined as
\begin{equation}
P(\sigma_i\vert\sigma_{<i}) = \frac{\sum_{\sigma_{>i}} e^{-\beta E(\sigma_i,\sigma_{<i})}}{\sum_{\sigma_i,\sigma_{>i}} e^{-\beta E(\sigma_i,\sigma_{<i})}}
= \frac{Z(\sigma_i,\sigma_{<i})}{\sum_{\sigma_i} Z(\sigma_i,\sigma_{<i})},
\label{eq:cp}
\end{equation}
where \(\beta=1/T\), \(\sigma_{<i}\) denotes the partial configuration of the first \(i-1\) spins, and \(\sigma_{>i}\) represents the remaining \(N-i\) spins. The quantity \(Z(\sigma_i,\sigma_{<i})\) is the local partition function constrained by the fixed partial configuration \(\{\sigma_{<i},\sigma_i\}\).

As emphasized in Refs.~\cite{cut,chen2025tensor}, a major advantage of TNMC is that the conditional probabilities in Eq.~\eqref{eq:cp} can be efficiently evaluated via tensor-network contractions. Even when tensor truncation introduces approximate errors, the sampled configurations still asymptotically converge to the exact Boltzmann distribution in Eq.~\eqref{eq:boltzman}.

\subsection{TNMC Method for the $J_1-J_2$ spin model with Free Boundaries}

When simulating classical spin systems with the TNMC method, we first express the partition function in the form of tensor networks. For convenience, we take open boundary conditions as an example to illustrate the transformation from the spin representation of the original \(J_1\)-\(J_2\) lattice partition function to the definition of tensors and the construction of tensor networks. Afterwards, we elaborate on the caching techniques and detailed procedures required for the sampling algorithm, with special emphasis on the left, right, upper and lower tensors involved in sampling operations.

Before detailing the tensor construction, we summarize the overall procedure. Three preprocessing steps are performed once and are not counted as Monte Carlo steps: (i) build the tensor network on the rotated lattice; (ii) cache the upper tensors $A'_{y,x}$ by contracting rows from top to bottom; (iii) cache the right tensors $A''_{y,x}$ by contracting columns from right to left. Each Monte Carlo sweep then proceeds row by row from top to bottom. Within a row, spins are sampled from left to right: at each site we contract the upper, left, right and lower caches to obtain the conditional probability $P(\sigma_i=\pm1|\sigma_{<i})$, draw $\sigma_i$ from this Bernoulli distribution, project the sampled value into the network via the projection vector $\nu$, and update the left and lower caches. After completing a row, the right caches for the next row are recomputed before proceeding. The following subsections elaborate each step.

\subsubsection{Definition of tensors and construction of tensor networks}
\begin{figure}[htbp]
\centering
\includegraphics[width=\linewidth]{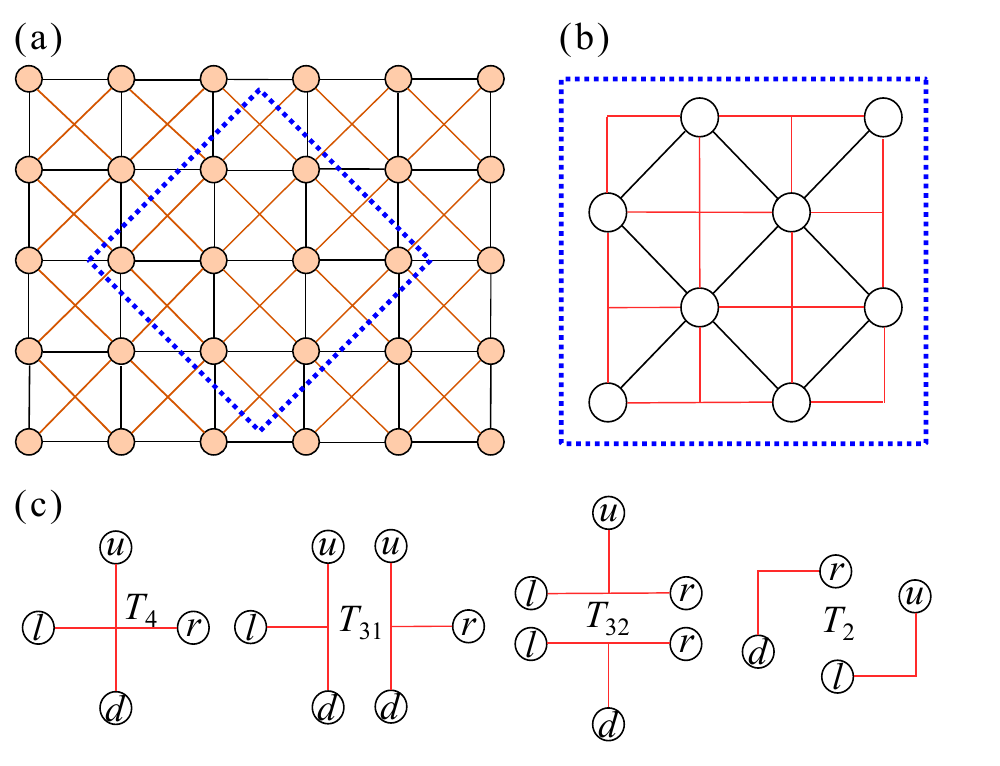}
\caption{Schematic diagram of the lattice structure and tensor decomposition for the two-dimensional square-lattice $J_1$-$J_2$ frustrated Ising model. (a) Spin representation of the full two-dimensional lattice: circles denote spins, solid black lines stand for nearest-neighbor couplings $J_1$, red diagonal lines represent next-nearest-neighbor diagonal couplings $J_2$, and the blue dashed box marks the lattice region to be transformed. (b) The grid enclosed by the blue dashed box in panel (a) is extracted and rotated clockwise by $45^\circ$, yielding the $4\times4$ tensor grid shown here; circles correspond to copy tensors, and intersections of red lines denote interaction tensors $T$. (c) Four types of fundamental local tensors $T_4$, $T_{31}$, $T_{32}$, and $T_2$, where the indices $l/r/u/d$ correspond to the left, right, upper, and lower spin connecting degrees of freedom, respectively.}
\label{fig:tn}
\end{figure}

The construction of tensor networks is not unique. Previous work~\cite{cut} performs singular value decomposition on the Boltzmann weight matrices defined over edges, and traces out the local physical spin degrees of freedom at each lattice site. The contributions from each site are then rearranged into local site tensors carrying auxiliary virtual indices. By subsequently connecting these tensors according to the lattice topology, the partition function of the entire system is recast as a tensor-network contraction problem. In contrast, Ref.~\cite{chen2025tensor} avoids decomposing Boltzmann matrices. Instead, its tensors are defined by absorbing two adjacent Boltzmann matrices into the copy tensor at each lattice vertex.
Neither of the above two strategies is well suited for the $J_1$-$J_2$ frustrated spin model. In this work, we adopt the scheme proposed in Refs.~\cite{li2021tensor,hotrgj1j2023}, where interaction tensors $T$ and copy tensors are directly constructed.

Figure~\ref{fig:tn}(a) displays a schematic of the $J_1$-$J_2$ Ising model with open boundary conditions. After extracting the lattice grid enclosed by the blue dashed box, we rotate it clockwise by $45^\circ$. This rotation is essential: after rotation, the diagonal $J_2$ bonds align with the horizontal and vertical virtual directions, so that copy tensors (one per Ising spin) and interaction tensors $T$ (one per plaquette) can alternate on a regular grid. We then construct the $4\times4$ tensor grid illustrated in Fig.~\ref{fig:tn}(b). This grid consists of alternating spin copy tensors $\delta$ (represented by open circles) and coupling interaction tensors $T$ (red crosses).
The copy tensor is a four-index Kronecker delta tensor, denoted by the hollow circles in Fig~\ref{fig:tn}(b), and is defined via the formula below:

\begin{equation}
\begin{aligned}
 & \delta_{\mathrm{lrud}}=
\begin{cases}
1 & \mathrm{~for~}l=r=u=d \\
0 & \text{ otherwise.}
\end{cases}
\end{aligned}
\label{eq:copy}
\end{equation}
Accordingly, the tensors located at the boundaries are defined as three-order $\delta$ tensors, while the tensors at the corners reduce to two-order $\delta$ tensors.
The system contains four types of local coupling tensors, namely $T_4$, $T_{31}$, $T_{32}$, and $T_2$, which correspond to bulk tensors, left/right boundary tensors, upper/lower boundary tensors, and corner tensors, respectively. These tensors are illustrated by the red grid elements in Fig.~\ref{fig:tn}(c). Taking the interaction tensor on the left boundary as an example, this tensor incorporates two nearest-neighbor interactions $\sigma_u\sigma_r$ and $\sigma_r\sigma_d$, as well as one next-nearest-neighbor interaction $\sigma_u\sigma_d$. Following the same rule, the explicit forms of the interaction tensors on all four boundaries can be obtained as follows:

\begin{equation}
\begin{aligned}
 & T_{\mathrm{rud}}=e^{(\beta/2)(\sigma_u\sigma_r+\sigma_r\sigma_d)-\beta g(\sigma_u\sigma_d)},\quad \text{left boundary}, \\[4pt]
 & T_{\mathrm{lud}}=e^{(\beta/2)(\sigma_l\sigma_u+\sigma_d\sigma_l)-\beta g(\sigma_u\sigma_d)},\quad \text{right boundary}, \\[4pt]
 & T_{\mathrm{lrd}}=e^{(\beta/2)(\sigma_r\sigma_d+\sigma_d\sigma_l)-\beta g(\sigma_l\sigma_r)},\quad \text{upper boundary}, \\[4pt]
 & T_{\mathrm{lru}}=e^{(\beta/2)(\sigma_l\sigma_u+\sigma_u\sigma_r)-\beta g(\sigma_l\sigma_r)}, \quad \text{lower boundary}.
\end{aligned}
\end{equation}
Each expression contains half-weighted nearest-neighbor coupling terms and a full next-nearest-neighbor coupling term, which completely characterizes the local interaction environment of boundary spins in the $J_1$-$J_2$ model.
We also include the corner tensors $T_{\mathrm{rd}}=e^{(\beta/2)(\sigma_r\sigma_d)}$ and $T_{\mathrm{lu}}=e^{(\beta/2)(\sigma_l\sigma_u)}$.

In practical programming implementations, all tensors are constructed as fourth-order tensors to facilitate tensor contraction, singular value decomposition (SVD), and other numerical operations. For tensors with fewer than four indices (i.e., second-order or third-order tensors), we augment their order by introducing redundant dimensions of size one. For instance, a second-order tensor can be reshaped into a fourth-order form $1 \times 1 \times d_1 \times d_2$. With all tensors cast in this uniform fourth-order form, the network can be contracted systematically; the caching scheme that avoids redundant contractions is presented in the next subsection.

\begin{figure}[t]
\centering
\includegraphics[width=\linewidth]{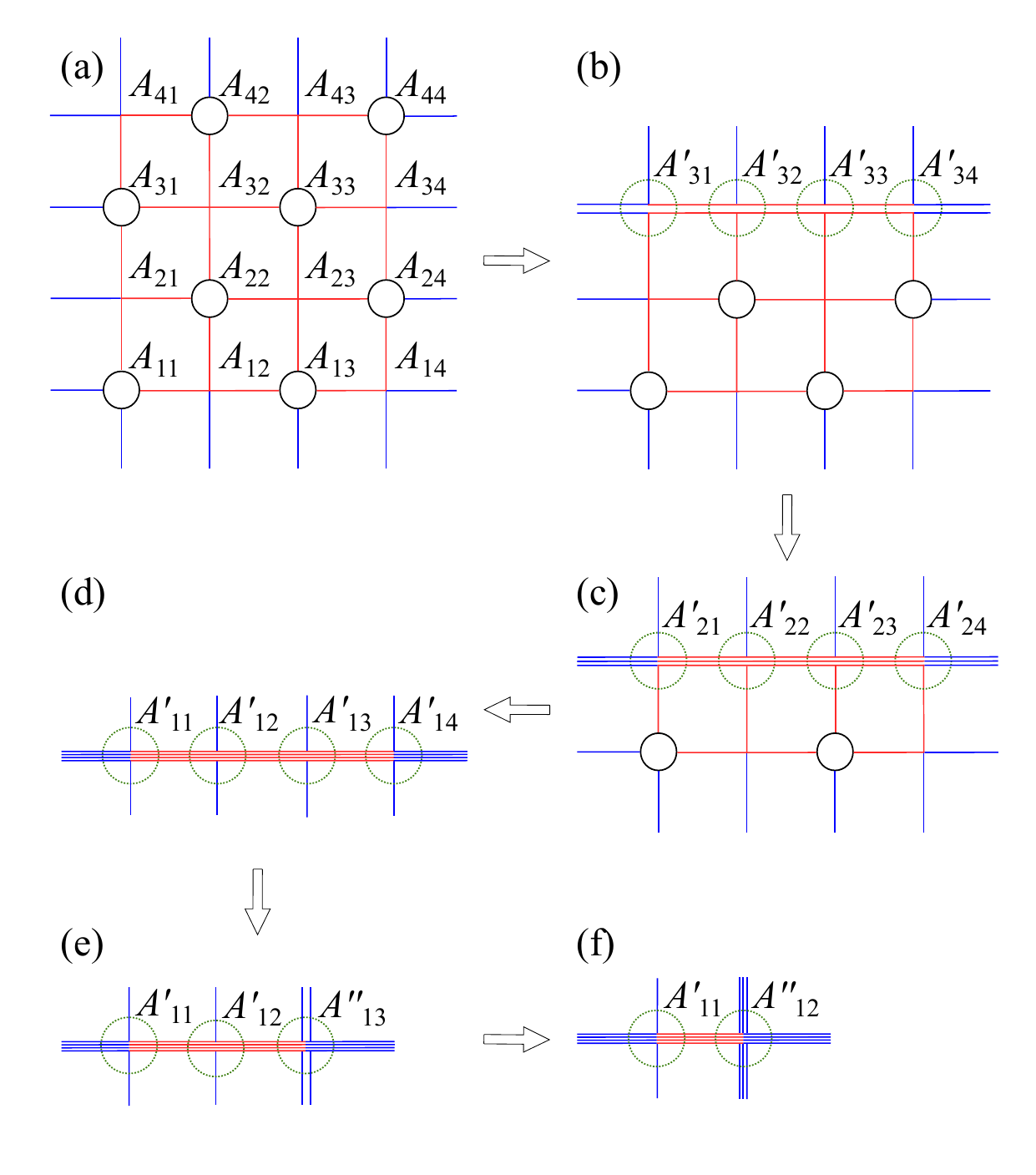}
\caption{Caching and contraction workflow for autoregressive sampling on the $4\times4$ tensor grid. Panel (a) shows the initial grid, while panels (b)--(d) depict the top-to-bottom row contraction procedure. The intermediate tensors $A'_{y,x}$ generated during this process are cached to avoid redundant computations. Finally, column contraction is performed from right to left on the bottom layer to extract the full partition function. Blue lines denote indices with bond size 1, and red lines denote indices with bond size 2. Note that all subfigures merely illustrate the topological order of caching and contraction and do not reflect the actual geometric dimensions of the tensors, as dimension reduction via QR and SVD is applied throughout the procedure.}
\label{fig:4x4suobing}
\end{figure}

\subsubsection{Top-to-bottom row-wise tensor network caching}
\label{sec:topdown}

To perform spin sampling, it is necessary to evaluate the conditional probability in Eq.~\ref{eq:cp}. Repeated tensor construction and contraction can be avoided by adopting a caching strategy.

Figure~\ref{fig:4x4suobing}(a) shows the schematic of the tensor network, where each tensor is denoted as $A_{y,x}$, with subscripts $y$ and $x$ representing the row and column indices, respectively. The caching procedure for tensor contraction proceeds in two sequential directions. First, the tensors are contracted row by row from top to bottom, followed by a right-to-left contraction along the bottom row. The cached quantities include the logarithm of the partition function and all intermediate tensors generated during contraction. 

For the topmost row, the initial matrix product state (MPS) tensors are identical to the original grid tensors, i.e., $A_{4,x}' = A_{4,x}$. By contracting these tensors $A_{4,x}'$ with the matrix product operator (MPO) tensors $A_{3,x}$ in the row directly below them, new intermediate tensors $A_{3,x}'$ are obtained, as illustrated in Fig.~\ref{fig:4x4suobing}(b). This procedure is repeated iteratively to generate $A_{2,x}'$ and $A_{1,x}'$, as shown in Fig.~\ref{fig:4x4suobing}(c) and Fig.~\ref{fig:4x4suobing}(d), completing the top-to-bottom caching of the intermediate tensor network $A_{y,x}'$.

Subsequently, the right-to-left caching is implemented. In Fig.~\ref{fig:4x4suobing}(d), $A_{14}'$ serves as the initial rightward cached tensor $A_{14}''$. Contracting $A_{14}''$ with $A_{13}'$ yields $A_{13}''$, and this process continues to produce $A_{12}''$ and $A_{11}''$. The final tensor $A_{11}''$ is a scalar value, which corresponds to the full partition function defined in Eq.~(2).

\begin{figure}[tbp]
\centering
\includegraphics[width=\linewidth]{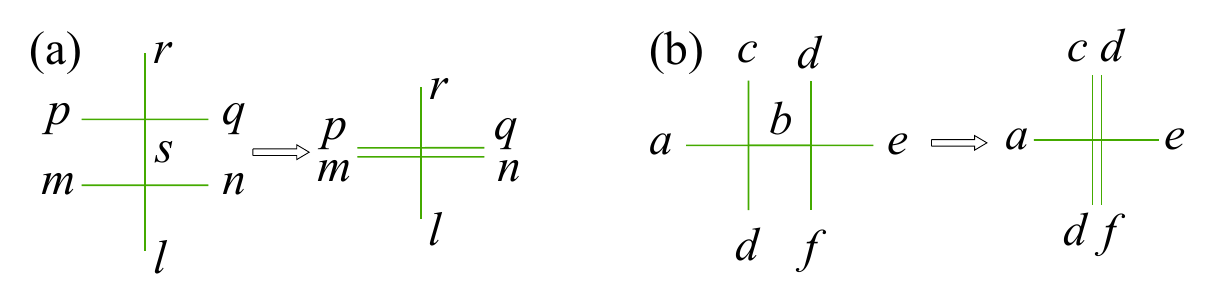}
\caption{ Rules for simplifying Einstein summation indices in TN contraction caching: (a) parallel index-line combination; (b) perpendicular index-line combination. }
\label{fig:suobingzhibiao}
\end{figure}

The full partition function above corresponds to the small \(4\times4\) grid; in practical simulations the tensor network is far larger, and direct contraction will trigger numerical overflow. For this reason, we apply local normalization when contracting boundary MPS/MPO row by row or column by column at each step. The logarithm of each normalization factor is added to \(\ln Z\).
In addition, Fig.~\ref{fig:4x4suobing} only shows the order of caching and contraction. It does not display the real sizes of tensor indices. In real calculations, we reduce index sizes via QR decomposition and SVD truncation~\cite{vidal2007itebd}. When we multiply an MPS tensor with an MPO tensor, the virtual bond dimension grows from \(D\) to \(Dd\). This causes fast expansion of tensor sizes. To avoid such expansion, the boundary MPS algorithm uses SVD truncation after each multiplication. This keeps the maximum bond dimension no larger than the preset cutoff value \(D\).

We now introduce the details of the top-down and right-left tensor contraction procedures. As shown in Fig.~\ref{fig:suobingzhibiao}(a), the cached boundary MPS tensor accumulated from the upper layer is denoted as $A'_{y+1,x,\mathit{pqrs}}$, and the local MPO tensor of the current layer is $A_{y,x,\mathit{mnsl}}$. These two tensors contract over the shared vertical physical index $s$ following the Einstein summation convention. The updated MPS tensor of the current layer $A'_{y,x,\mathit{pmqnrl}}$ is given by
\begin{equation}\label{eq:downward_middle}
    A'_{y,x,\mathit{pmqnrl}} = \sum_{s}  A'_{y+1,x,\mathit{pqrs}} \cdot A_{y,x,\mathit{mnsl}}
\end{equation}
where the summation runs over the shared index $s$. The resulting six-index tensor is reshaped as four-index tensor, and truncated via singular value decomposition SVD and QR, and is finally converted back into a standard four-index tensor form. 
The above contraction rule applies to general middle layers.
For the bottom layer ($y=1$, the first row), the extra upper index $r$ and lower index $s$ are further contracted:
\begin{equation}\label{eq:downward_bottom}
    A'_{1,x,\mathit{pmqn}} = \sum_{r} \sum_{s}  A'_{2,x,\mathit{pqrs}} \cdot A_{1,x,\mathit{mnsr}}
\end{equation}

Figure~\ref{fig:suobingzhibiao}(b) illustrates the details of the right-to-left contraction. The valid cached tensor on the right side is $A''_{y,x+1,\mathit{bedf}}$, which is accumulated from previous rightward contraction. The cached tensor of the current column is $A'_{y,x,\mathit{abcd}}$. These two tensors contract over the shared index $b$, and the updated right-side valid tensor $A''_{y,x,\mathit{aecdbf}}$ satisfies
\begin{equation}\label{eq:leftward_contraction}
    A''_{y,x,\mathit{aecdbf}} = \sum_{b} A''_{y,x+1,\mathit{bedf}} \cdot A'_{y,x,\mathit{abcd}}
\end{equation}
The tensor obtained from this right-to-left contraction is also reshaped and converted into a standard four-index tensor.

In summary, our contraction caching scheme stores all intermediate tensors and the accumulated $\ln Z$ from each row.

\subsubsection{Spin Projection for Sampling}
\label{sec:projection}

To compute the spin sampling probability $P(\sigma_i|\sigma_{<i})$ at each lattice site in Eq.~\ref{eq:cp}, a spin projection operation is required. For Ising spins, we evaluate $P(\sigma_i=1|\sigma_{<i})$ and $P(\sigma_i=-1|\sigma_{<i})$ for $\sigma_i=\pm1$. In the tensor network framework, a simple approach is to replace the original copy tensor in Eq.~\ref{eq:copy} with a one-dimensional projection vector $\nu_a$, defined as

\begin{equation}\label{eq:spin_vector}
    \nu_a =
    \begin{cases}
    \delta_{a,1} = (0,\,1)^\top, & \sigma_i = +1\\
    \delta_{a,2} = (1,\,0)^\top, & \sigma_i = -1
    \end{cases}
\end{equation}

\begin{figure}[b]
\centering
\includegraphics[width=\linewidth]{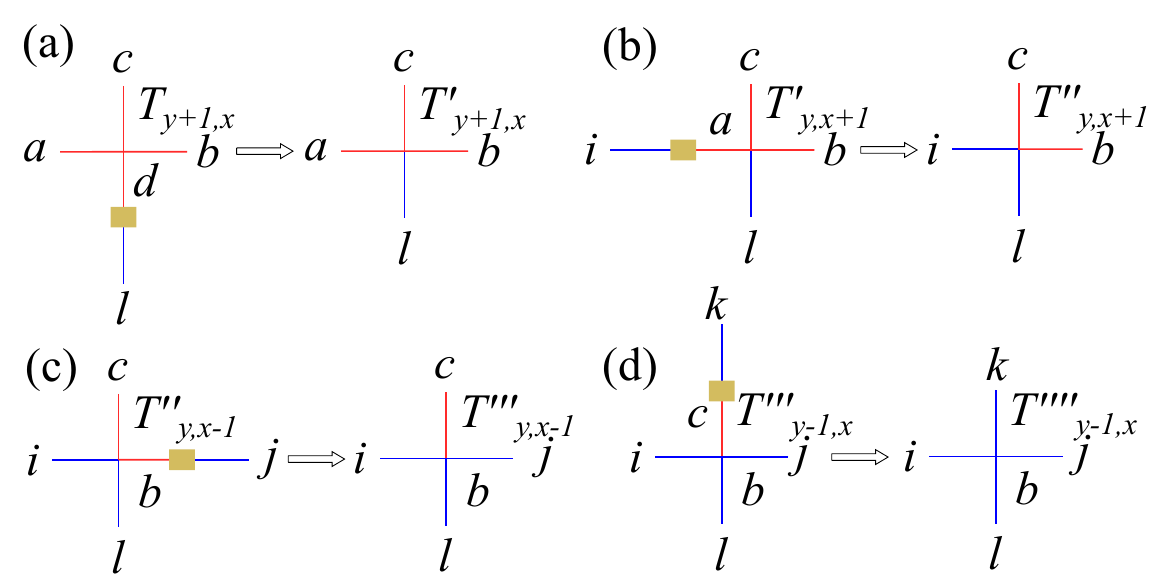}
\caption{Update process of a single fourth-order interaction tensor. The yellow squares denote the projection vector $\nu_a$.}
\label{Fig:cp_T}
\end{figure}
After the spin value at site $(y,x)$ is determined, the original copy tensor is replaced by the projection tensor $\nu$. 
 This projection tensor is further contracted with the connected interaction tensor at $(y+1,x)$ to evaluate the corresponding partial partition function.

Figure~\ref{Fig:cp_T} shows several typical contraction cases of projection tensors and interaction tensors.
In Fig.~\ref{Fig:cp_T}(a), the interaction tensor $T_{abcd}$ is located at the site $(y+1,x)$. Its index $d$ corresponds to the sampled spin degree of freedom $\sigma_i$ at site $(y,x)$, while the remaining indices $a$, $b$, and $c$ correspond to other unsampled neighboring spins. Red lines represent indices with a bond dimension of 2. The yellow square at site $(y,x)$ indicates that the spin $\sigma_i$ is assigned a value of $+1$ or $-1$. The projection tensor is one-dimensional, marked by blue lines. To embed the constraint of the sampled spin into the interaction tensor $T_{abcd}$, we perform tensor contraction on the shared index $d$. The specific form is given by:

\begin{equation}\label{eq:touyingd}
    T'_{abcl} = \sum_d T_{abcd} \, \nu_{dl}.
\end{equation}

As shown in Fig.~\ref{Fig:cp_T}(b), the four-leg interaction tensor at $(y,x+1)$ contracts with the projection tensor at $(y,x)$. The bond dimension of its bottom leg reduces to 1, marked by the blue line in the figure. However, the bond dimension of its left leg remains 2, so we contract this left leg with the projection tensor at $(y,x)$. The tensor update formula reads:

\begin{equation}\label{eq:touyinga}
    T''_{ibcl} = \sum_{a} T'_{abcl} \, \nu_{ai}.
\end{equation}

As shown in Fig.~\ref{Fig:cp_T}(c), the projection tensor at $(y,x)$ contracts with the interaction tensor on its left at $(y,x+1)$. The corresponding formula is
\begin{equation}\label{eq:touyingb}
    T'''_{ijcl} = \sum_{b} T''_{ibcl} \, \nu_{bj}
\end{equation}
Before contraction, the bond dimensions of its bottom and left legs (the 1st and 4th indices) are reduced to 1, which are labeled by blue lines.
Similarly, Fig.~\ref{Fig:cp_T}(d) describes the contraction between the projection tensor at $(y,x)$ and the tensor at $(y-1,x)$. At this stage, the bottom, left and right legs (the 1st, 2nd and 4th indices) of this tensor all have bond dimension 1, indicated by blue lines. The update formula for this contraction reads
\begin{equation}\label{eq:touyingc}
    T''''_{ijkl} = \sum_{c} T'''_{ijcl} \, \nu_{ck}
\end{equation}
\subsubsection{Left, Right and Upper Tensors}

\begin{figure}[t]
\centering
\includegraphics[width=\linewidth]{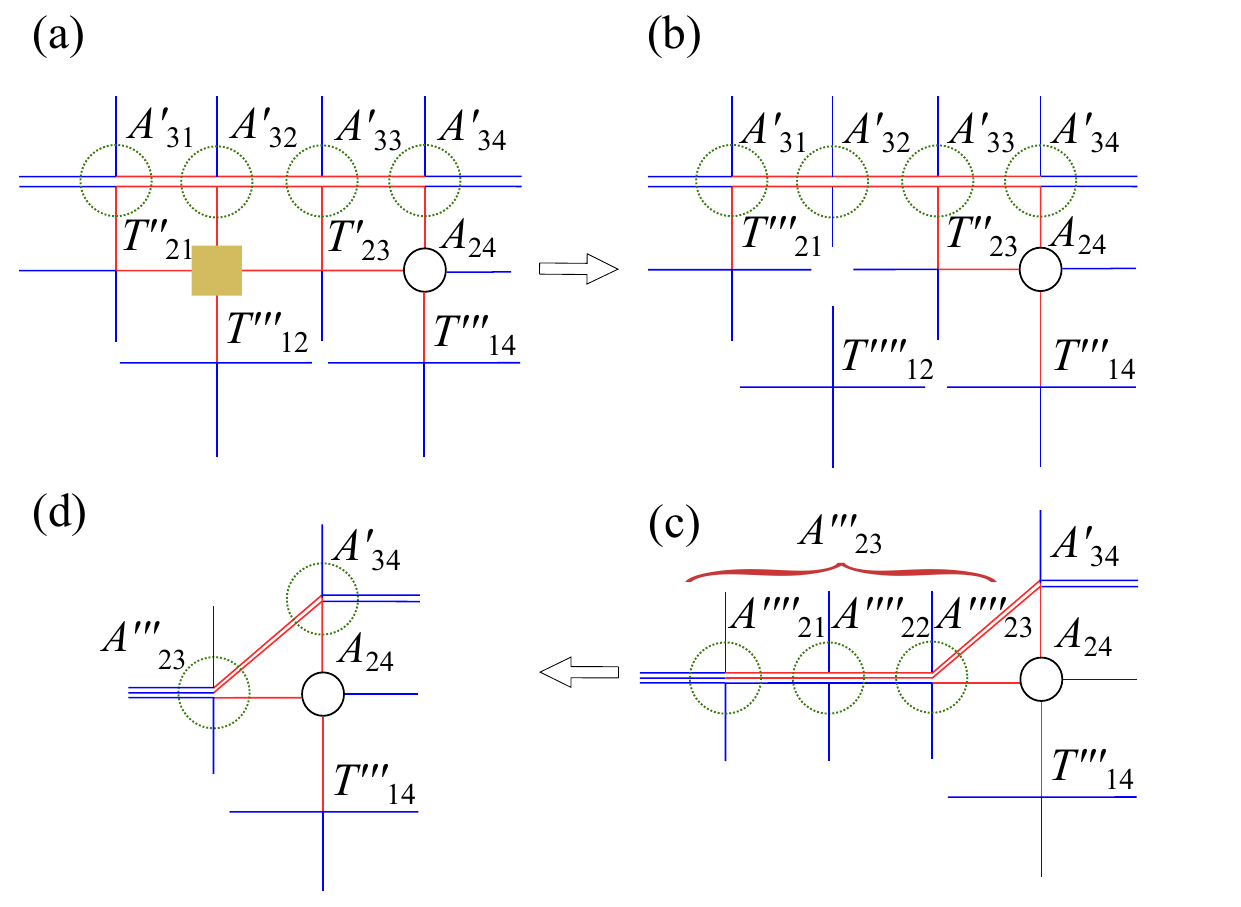}
\caption{Definition and caching schematic of left tensors (a) After sampling the spin at lattice site \((2,2)\) (yellow square), projections are applied to the four adjacent tensors. (b)  After projection, the index dimensions connecting the four neighboring tensors to the yellow square shrink to 1, where the red lines turn into blue lines.
(c) All tensors with \(x<4\) are combined into the left tensor \(A_{23}'''\), and the standalone tensor \(T_{12}''''\) is multiplied into the partition function.
(d) Take \(A_{24}\) as the central tensor, with \(A_{23}'''\) serving as the left tensor, \(T_{14}'''\) the lower tensor, and \(A_{34}'\) the upper tensor.}
\label{fig:leftsuobing}
\end{figure}

When we sample spin $\sigma_i$, four types of tensors are used in calculations: upper tensor, left tensor, right tensor and lower tensor. Their notations and physical meanings are listed below:
\begin{enumerate}
    \item Upper tensor (written as $\mathrm{A}'$): Contraction results cached by top-to-bottom tensor summation.
    Take the spin at the yellow block $(y,x)=(2,2)$ in Fig.~\ref{fig:leftsuobing}(a) as an example. Its upper tensor is $\mathrm{A}_{32}'$.
    \item Right tensor (written as $\mathrm{A}''$): Contraction outputs cached by right-to-left iteration.
    For the spin at $(y,x)=(2,2)$ in Fig.~\ref{fig:leftsuobing}(a), its full right tensor consists of $\mathrm{A}_{33}'$, $\mathrm{A}_{34}'$, $\mathrm{A}_{24}$, $\mathrm{T}_{23}'$ and $\mathrm{T}_{14}'''$.
    \item Left tensor (written as $\mathrm{A}'''$): Calculation outputs accumulated from left-to-right iteration.
    Look at the spin at $(y,x)=(2,4)$ on the right side of the yellow block in Fig.~\ref{fig:leftsuobing}(b). Its full left tensor includes $\mathrm{A}_{31}'$, $\mathrm{A}_{32}'$, $\mathrm{A}_{33}'$, $\mathrm{T}_{21}'''$ and $\mathrm{T}_{23}''$. These separate left tensors are further contracted into one unified left tensor $\mathrm{A}_{23}'''$, as shown in Fig.~\ref{fig:leftsuobing}(c) and (d).
    \item Lower tensor (written as $\mathrm{T}'''$): Determined by interaction constraints from all previously sampled fixed spins in the same row. It carries correlation information of sampled spins to the current lattice site. For example, the lower tensor at lattice site \((2,4)\) is denoted as \(T_{14}'''\).
\end{enumerate}

At each site $(y,x)$, the conditional probability required in Eq.~\eqref{eq:cp} is obtained by contracting the upper tensor $A'_{y+1,x}$, the left tensor $A'''_{y,x-1}$, the right tensor $A''_{y,x+1}$, and the lower tensor $T'''_{y,x}$ with the local copy tensor. This leaves a two-dimensional vector over $\sigma_i=\pm1$; normalizing it gives $P(\sigma_i=+1|\sigma_{<i})$ and $P(\sigma_i=-1|\sigma_{<i})$, from which the spin is drawn. There is a fixed order for these tensor calculations: the upper, right and lower tensors are fully computed before sampling each row, while the left tensor stores contributions from the previous lattice site and is updated between consecutive spins within a row. Among the four caches, the upper and lower tensors are already constructed in Secs.~\ref{sec:topdown} and~\ref{sec:projection}; in the following we focus on the left and right caches, which require careful bookkeeping during the row-wise sweep.

\paragraph{Left tensor}

After sampling the spin at site $(2,2)$, we save and process a set of tensors. These tensors act as the left tensors required for sampling the spin at $(y=2,x=4)$.

\begin{figure}[t]
\centering
\includegraphics[width=\linewidth]{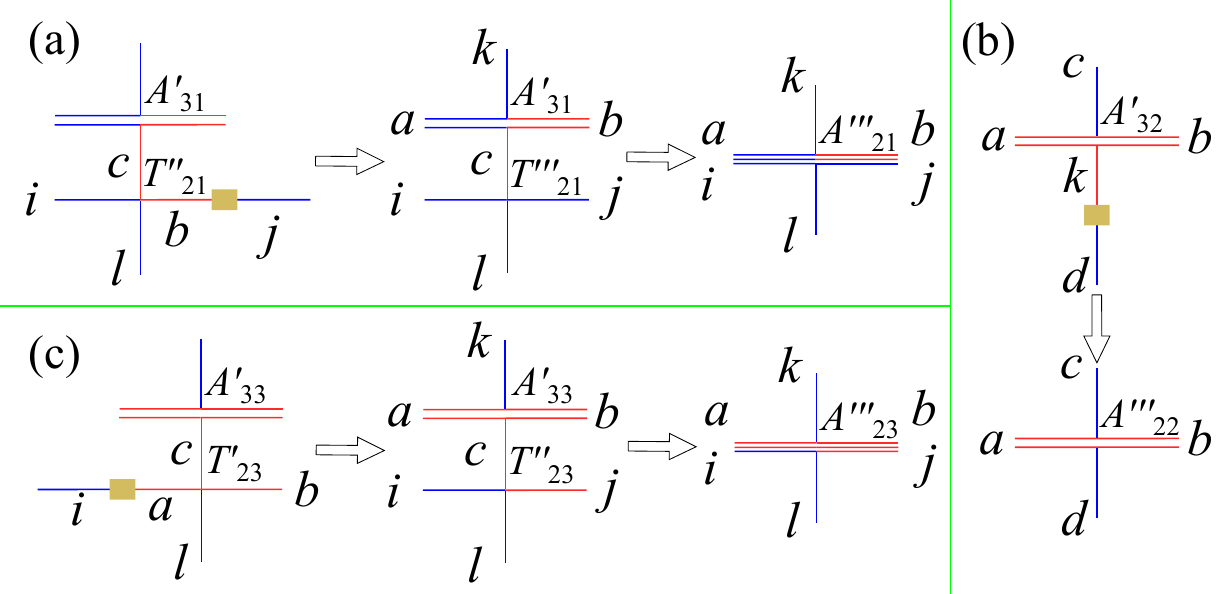}
\caption{Schematic of Contraction and Calculation Process for Left Tensors. The projected tensor's yellow square is on the right (a), below (b), and on the left (c).}
\label{fig:zuozhibiao}
\end{figure}

As shown in Fig.~\ref{fig:zuozhibiao}(a), after sampling the spin at $(2,2)$, we send the projection vector $\nu$ marked by the yellow block to its neighboring interaction tensor $T_{21}''$. We sum over the shared index $b$ following Eq.~\ref{eq:touyingb} to get $T_{21}'''$. Then we contract $T_{21}'''$ with $A_{31}'$ to get the intermediate tensor $A_{21}''''$, which is written as
\begin{equation}\label{eq:left_cache_21}
    A''''_{21,\mathit{aibjkl}} = \sum_{c} A'_{31,\mathit{abkc}} \, T'''_{2, 1,\mathit{ijcl}}
\end{equation}
In Fig.~\ref{fig:zuozhibiao}(b), the fixed spin value at $(2,2)$ affects the upper tensor $A'_{32, \mathit{abck}}$ used in sampling. We contract this upper tensor with vector $\nu_{22,\mathit{kd}}$ and sum over the shared index $k$ to obtain the intermediate tensor $A''''_{22,\mathit{abcd}}$. The formula is
\begin{equation}\label{eq:left_cache_22}
    A''''_{22,\mathit{abcd}} = \sum_{k} A'_{32,\mathit{abck}} \, \nu_{22,\mathit{kd}}
\end{equation}

As shown in Fig.~\ref{fig:zuozhibiao}(c), after sampling the spin at $(2,2)$, the projection vector is passed to the neighboring interaction tensor $T_{23}'$. We sum over the shared index $a$ according to Eq.~\ref{eq:touyinga} to get $T_{23}''$. Next, we contract $A'_{33,\mathit{abkc}}$ with $T''_{23,\mathit{ijcl}}$ and obtain the intermediate tensor $A''''_{23,\mathit{aibjkl}}$. The formula is
\begin{equation}\label{eq:left_cache_23}
    A''''_{23,\mathit{aibjkl}} = \sum_{c} A'_{33,\mathit{abkc}} \, T''_{23,\mathit{ijcl}}
\end{equation}

We now have three intermediate tensors $A''''_{21,\mathit{abcd}}$, $A''''_{22,\mathit{cdef}}$ and $A''''_{23,\mathit{efgh}}$ as drawn in Fig.~\ref{fig:leftsuobing}(c). We contract all three tensors together to produce the left cached tensor $A'''_{23,\mathit{abgh}}$ for sampling the spin at $(2,4)$. The corresponding formula reads
\begin{equation}\label{eq:left_cache_merge}
    A'''_{23,\mathit{abgh}} = \sum_{c} \sum_{d} \sum_{e} \sum_{f} A''''_{21,\mathit{abcd}} \cdot A''''_{22,\mathit{cdef}} \cdot A''''_{23,\mathit{efgh}}
\end{equation}

This final result $A_{23}'''$ serves as the left cached tensor for site $(2,4)$. After this step, we start sampling the spin at $(2,4)$.

\paragraph{ Right tensor}

\begin{figure}[htbp]
\centering
\includegraphics[width=\linewidth]{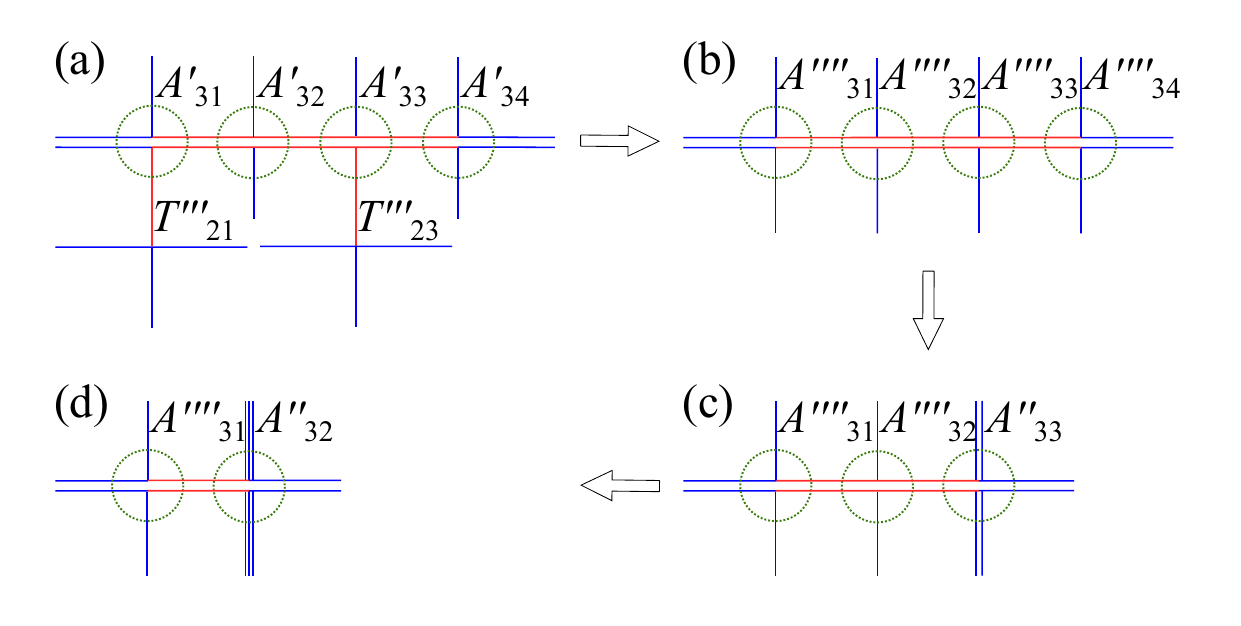}
\caption{Contraction and caching schematic of right tensors for the tensor at (3,1). (a) Configuration after sampling site \((2,2)\) and contracting the projected tensor $v$  at (2,2). (b) Tensors at \(y=2\) are also contracted into the tensor at \(y=3\).
 (c) \(A_{34}'''\) is contracted with \(A_{33}'''\) to form \(A_{33}''\), which serves as the right tensor at lattice site \((3,2)\). (d) \(A_{33}''\) is contracted with \(A_{32}''''\) to form \(A_{32}''\), which serves as the right tensor at lattice site \((3,1)\).}
\label{fig:rightsuobing}
\end{figure}
As shown in Fig.~\ref{fig:rightsuobing}(a), we finish sampling the spins at sites $(2,2)$ and $(2,4)$, and then start sampling the spin at $(3,1)$. All spins in the second row have been sampled, and their projection vectors have been sent to the interaction tensors. Before sampling spins in the third row, we compute the right cached tensors through a right-to-left sweep.

Figure~\ref{fig:rightsuobing}(a) plots the tensor network after all spins in the second row are fixed. The upper cached tensors $A_{31}'$, $A_{32}'$, $A_{33}'$, $A_{34}'$ connect with the interaction tensors $T_{21}'''$ and $T_{23}'''$. We contract $T_{21}'''$ and $T_{23}'''$ into their matching upper cached tensors $A_{31}'$ and $A_{33}'$ separately. This produces updated intermediate tensors $A_{31}''''$, $A_{32}''''$, $A_{33}''''$, $A_{34}''''$, as displayed in Fig.~\ref{fig:rightsuobing}(b).

Afterwards, we merge adjacent intermediate tensors one by one, focusing on the tensors circled by green dashed lines:
Contract $A_{33}''''$ and $A_{34}''''$ to get $A_{33}''$ shown in Fig.~\ref{fig:rightsuobing}(c);
Contract $A_{32}''''$ and $A_{33}''$ to get $A_{32}''$ shown in Fig.~\ref{fig:rightsuobing}(d).
For the spin at $(3,1)$, its right cached tensor is $A_{32}''$. For the spin at $(3,3)$, its right cached tensor is $A_{34}''''$. All right cached tensors needed for sampling spins in the third row are fully calculated after these steps.

\begin{figure}[b]
\centering
\includegraphics[width=\linewidth]{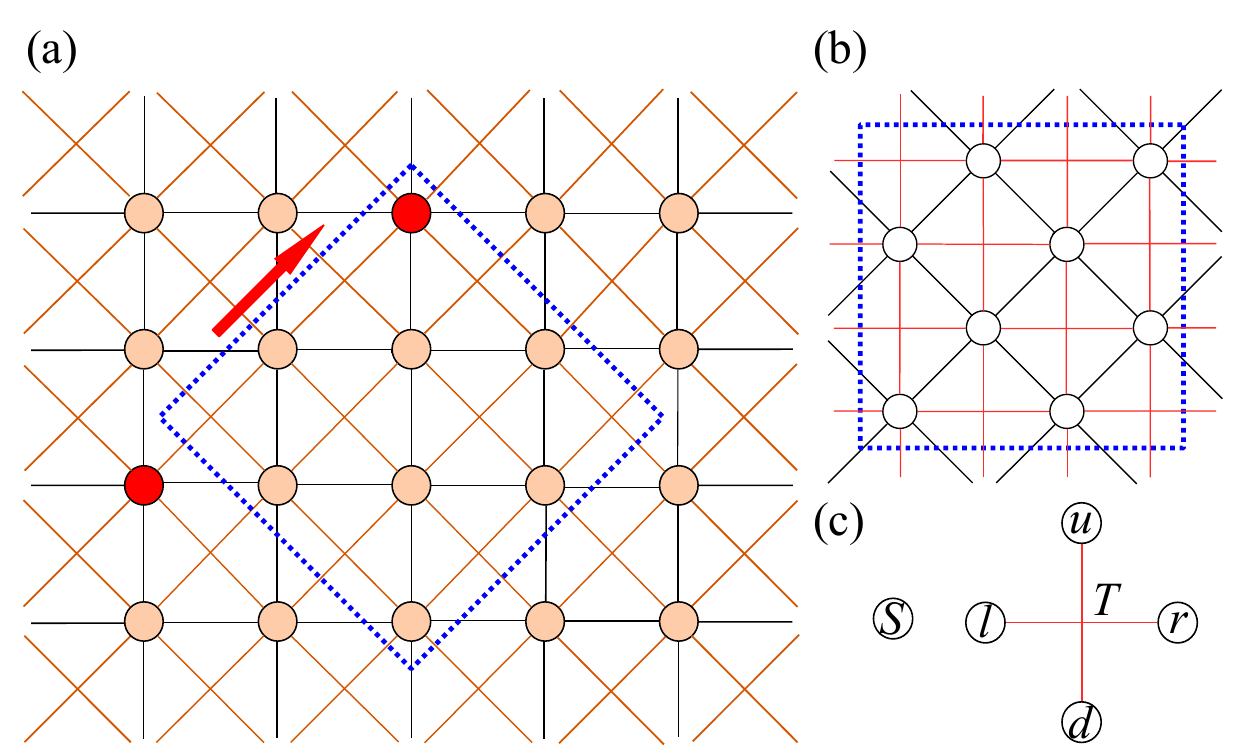} 
\caption{Tensor construction schematic under periodic boundary conditions. (a) Original lattice: spins inside the blue dashed box, arrows denote periodic boundary directions, red circles label equivalent spins under periodic boundaries. (b) Tensor network after 45$^\circ$ rotation of the left dashed lattice; circles are spin copy tensors, red crosses are interaction tensors. (c) Copy tensors and interaction tensor.}
\label{fig:zhouqi_struct}
\end{figure}

\subsection{Periodic-boundary lattice tensor networks}
\label{sec:pbc}

The sampling procedure described above is for open boundaries. We now extend the tensor-network construction to periodic boundary conditions. To eliminate boundary effects and make our physical model reproduce realistic bulk properties, we perform all calculations with periodic boundary conditions. Figure~\ref{fig:zhouqi_struct}(a) displays the lattice geometry of the J1-J2 Ising model under periodic boundaries. The finite simulation domain is enclosed by a blue dashed box, red arrows indicate the directions of periodic boundaries, and the two red circles mark two equivalent spin sites connected by periodicity.
For easier numerical simulation, we rotate the lattice region inside the blue dashed box in Fig.~\ref{fig:zhouqi_struct}(a) clockwise by 45 degrees to obtain the 4$\times$4 tensor network in Fig.~\ref{fig:zhouqi_struct}(b). This network contains two types of alternating tensors: spin tensors (circular copy tensors placed at each spin site) and four-leg interaction tensors (red cross-shaped $T$ tensors).
The explicit expressions for $T$ and $S$ tensors are written as
\begin{equation}
\begin{aligned}
 & T_{\mathrm{lrud}}=e^{(\beta/2)(\sigma_l\sigma_u+\sigma_u\sigma_r+\sigma_r\sigma_d+\sigma_d\sigma_l)-\beta g(\sigma_l\sigma_r+\sigma_u\sigma_d)}, \\
 & S_{\mathrm{lrud}}=\delta_{\mathrm{lrud}}=
\begin{cases}
1 & \mathrm{~for~}l=r=u=d \\
0 & \text{ otherwise.}
\end{cases}
\end{aligned}
\end{equation}

Preliminary numerical tests demonstrate that standard periodic boundary conditions lead to strong bond-dimension sensitivity for our \(J_1\)-\(J_2\) frustrated Ising model. In this test scheme, all boundary tensors are simply replaced by bulk tensors, which only enforces lattice periodicity geometrically but fails to achieve left-right canonicalization of the MPS without a well-defined canonical center. This greatly amplifies truncation errors during MPS contraction. Even a small reduction in the bond dimension \(D\) induces severe deviations in observables including partition functions and spin correlation functions. On the other hand, adopting a large \(D\)  to mitigate errors drastically increases memory usage and computational time, making high-precision simulations for large lattices intractable. To balance numerical accuracy and computational cost, we employ a block-sweeping scheme as an approximation to replace standard periodic boundaries.

\begin{figure}[t]
\centering
\includegraphics[width=\linewidth]{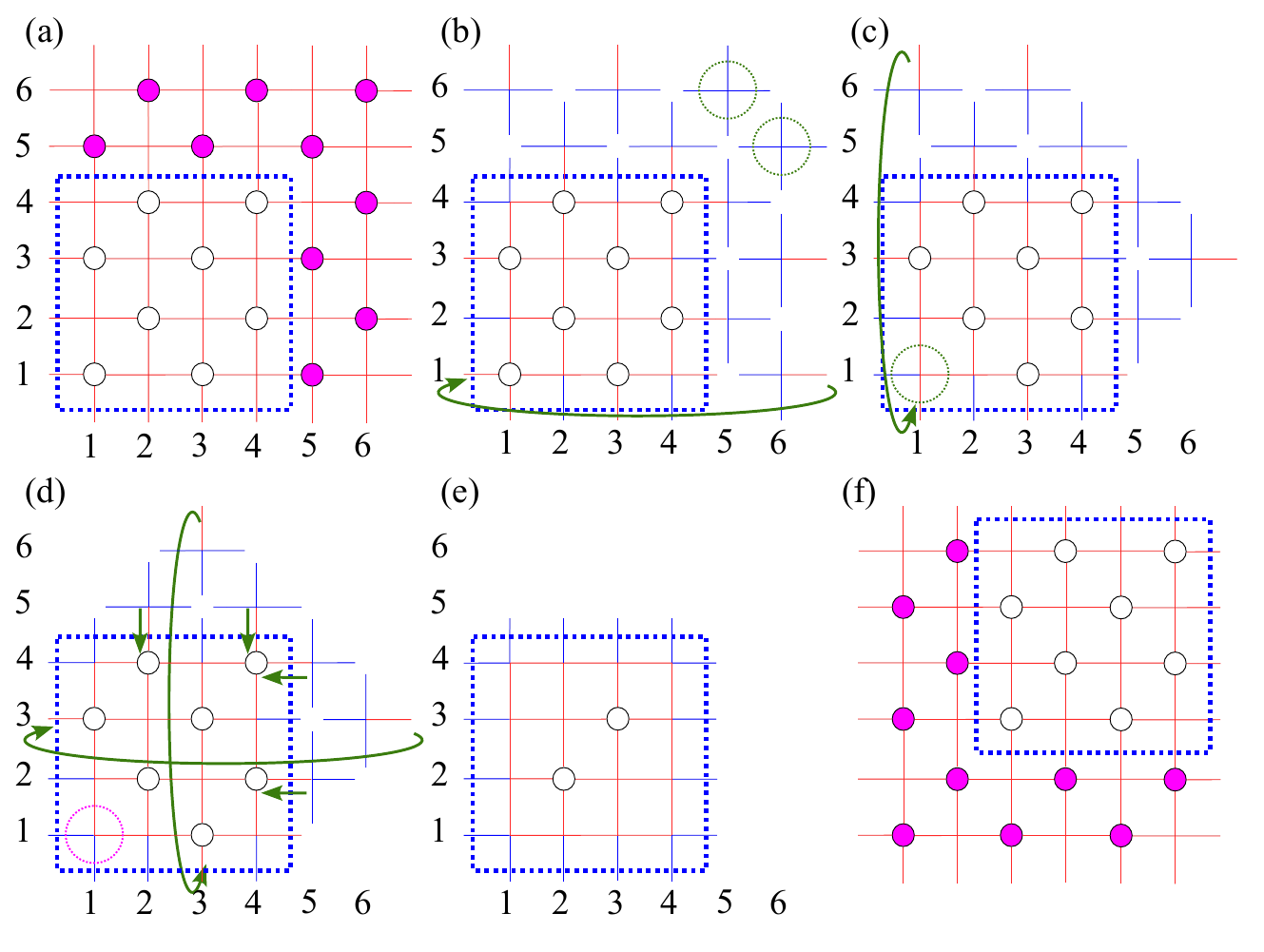}
\caption{Schematic for selecting fixed boundary spins and spins awaiting sampling in periodic-boundary lattice tensor networks.
(a) White circles: spins to be updated and sampled; pink spheres: fixed spins inherited from the prior configuration.
(b) Contraction outcome between projected tensors of fixed spins and neighboring interaction tensors; tensors inside the green dashed box are scalars.
Contraction along green arrows from (b) to (d) generates the quasi-open-boundary tensor network (e) for caching and sampling; red bars stand for dimension-2 indices, purple bars for dimension-1 indices.
(f) Another sampling region selection method, which can be transformed into a tensor network similar to panel (e).}
\label{fig:zhendang}
\end{figure}

Figures \ref{fig:zhendang}(a) and (f) present schematic illustrations of the block-sweeping boundary scheme. Purple circles stand for spins with fixed magnitudes, and white circles within dashed boxes correspond to spins to be sampled. For these fixed boundary spins, only the initialization stage employs the precomputed ground-state configuration; in all subsequent iteration steps, the spin configuration obtained from the previous sampling cycle is adopted as the initial state of the purple boundary spins in panel (f). The boundary width is set to 2 for the following reason: if the width were chosen as 1, the spins at lattice coordinates (1,1) and (5,5) would be vertically coupled to each other. This would inevitably require an excessively large bond dimension to compute partition functions or partial partition functions with acceptable precision. This consequently leads to a low sample acceptance probability as shown in Eq.~\ref{eq:pa}.

Subsequently, the boundary spins are fixed, so their corresponding projection tensors should be contracted into the neighboring interaction tensors. The full operational details are illustrated in Figs. \ref{fig:zhendang} (b)-(e).
Take the horizontal lattice row at \(y=6\) as an example, which contains three interaction tensors in total. After tensor contraction, the two left tensors each retain one red index leg and three blue index legs. The red indices couple to the unsampled spins on the \(y=1\) row and maintain a dimension of 2, whereas the three blue legs arise from the contraction of projection tensors for fixed boundary spins. There exists an additional interaction tensor at lattice site \((6,5)\), which is enclosed by the green dashed circle. After contraction, all four index legs of this tensor turn blue with the dimension reduced to 1, yielding a standalone scalar. This scalar only needs to be multiplied into the total partition function and does not take part in any subsequent tensor contraction procedures.

The subsequent task is to contract all tensors outside the blue dashed box into the region enclosed by the blue box. The detailed specifications are provided below. As shown in Fig.~\ref{fig:zhendang}(b), take the interaction tensor $T$ at site \((1,6)\) as an example. Following the direction indicated by the green arrow, the right leg of tensor $T$ at \((1,6)\) connects to the left leg of the spin-copy tensor at site \((1,1)\). Contract these two tensors to yield the new tensor enclosed by the green dashed circle in Fig.~\ref{fig:zhendang}(c). Apart from the horizontal direction, along the vertical direction in Figure \ref{fig:zhendang}(c), the lower leg of the spin-copy tensor at site \((1,1)\) connects to the upper leg of the interaction tensor at site \((6,1)\), and an identical contraction operation is carried out likewise.

Similarly, as shown in Fig~\ref{fig:zhendang}(d)  the right leg of tensor T at site \((3,6)\) is connected to the left leg of the copy tensor at site \((1,6)\), and the identical contraction operation is performed.
Along the vertical direction, the lower leg of the copy tensor at site \((1,3)\) is connected to the upper leg of the tensor at site \((6,3)\). Contraction is carried out following the direction indicated by the green arrow to further eliminate tensors outside the blue dashed box.
In addition, the two boundary tensors at \(y=5\) and the two boundary tensors at \(x=5\) are also contracted into the region enclosed by the blue dashed box following the directions marked by the green arrows.
Finally, the resulting tensor network exhibits a structure similar to the open-boundary case, illustrated in Fig.~\ref{fig:zhendang}(e). 
At this stage, only the two original copy tensors, denoted by hollow circles, remain inside the box.

Subsequently, we sample the spins inside the box following the open-boundary scheme introduced in the previous subsection. After this sampling step, we proceed to perform a swap update for the fixed-boundary configuration illustrated in Fig.~\ref{fig:zhendang}(f). Nevertheless, four lattice sites never get updated throughout the above procedure, namely the sites at coordinates \((y,x)=(5,1)\), \((6,2)\), \((1,5)\), and \((2,6)\). To resolve this issue, an additional Metropolis sweep is implemented prior to the boundary swap step, which attempts to flip every spin site over the entire lattice.

\begin{figure}[b]
\centering
\includegraphics[width=0.9\linewidth]{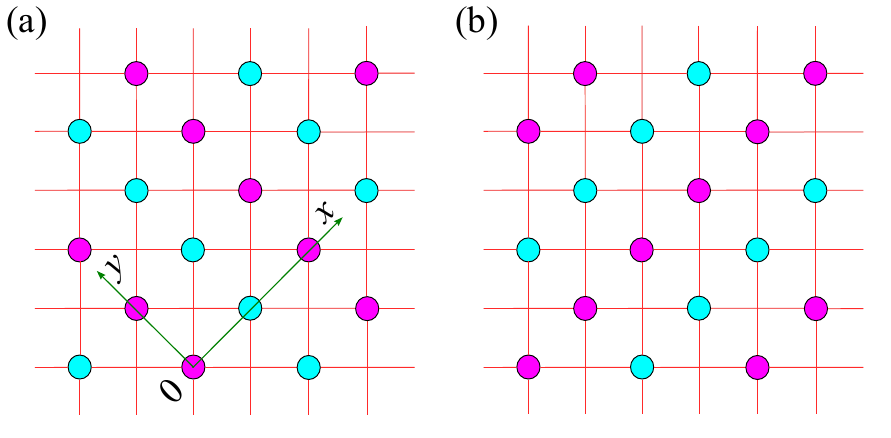}
\caption{Two typical stripe-phase configurations on the $6\times6$ lattice; the coordinate axes serve as the reference frame for measuring stripe magnetization and susceptibility later.}
\label{fig:st}
\end{figure}

\subsection{Observables}

\begin{itemize}
    \item The correlation time defined as~\cite{Sandvik_2010},
    \begin{equation}
        A_Q(t)=\frac{\langle Q_kQ_{k+t}\rangle-\langle Q_k\rangle^2}{\langle Q_k^2\rangle-\langle Q_k\rangle^2}.
    \end{equation}

    \item {Acceptance probability and effective sampling time}
    The acceptance probability is given by~\cite{chen2025tensor}
    \begin{equation}
        p_a=N_a/N_t,
    \end{equation} 
    where $N_a$ is the number of accepted moves and $N_t$ is the total number of trial moves. The effective sampling time is defined as~\cite{chen2025tensor} 
     \begin{equation}
      t_a=t_{\text{real}}/p_a,
      \label{eq:time}
    \end{equation}

    \item {System energy and specific heat capacity.}
    The system energy is calculated as $E=\langle H\rangle/N$. The specific heat capacity is formulated as
    \begin{equation}\label{eq:heat_capacity}
        C_V = \frac{1}{N T^2} \left( \langle E^2 \rangle - \langle E \rangle^2 \right)
    \end{equation}
    where $N$ denotes the number of spin particles in the system and $T$ is the thermodynamic temperature of the system.

    \item {Stripe-phase magnetization.}
    For $g>0.5$, the model enters the stripe phase, whose schematic diagram is illustrated in Fig.~\ref{fig:st}.
 The magnetization components for the stripe phase are given by
    \begin{equation}\label{eq:magnetization_xy}
        \begin{aligned}
            m_x &= \frac{1}{N} \sum_{i=1}^{N} \sigma_i (-1)^{x_i}, \\
            m_y &= \frac{1}{N} \sum_{i=1}^{N} \sigma_i (-1)^{y_i},\\
            m &= \sqrt{m_x^2 + m_y^2}
        \end{aligned}
    \end{equation}

    \item {Stripe susceptibility.}
   The stripe susceptibility is formulated as
    \begin{equation}\label{eq:stripe_susceptibility}
        \chi = \frac{N}{T} \left( \langle m^2 \rangle - \langle |m| \rangle^2 \right).
    \end{equation}

    \item {Binder cumulant}
    The Binder cumulant $Q$ is defined as
    \begin{equation}\label{eq:binder_q}
        Q= \frac{\langle m^4 \rangle}{\langle m^2 \rangle^2}
    \end{equation}
    where $m$ is the stripe order parameter defined in Eq.~\eqref{eq:magnetization_xy}, not the standard ferromagnetic magnetization. It is adopted to determine the critical point of the system. The curves of the Binder cumulant $Q$ versus temperature $T$ for systems with different finite sizes $L$ intersect at the critical temperature $T_c$ in the thermodynamic limit.
\end{itemize}

The complete pseudo-code of the TNMC Markov chain, including the block update, boundary swap, and Metropolis sweep, is summarized in Appendix~\ref{app:tnmc_algo}.

\section{Histogram Reweighting for the $J_1$-$J_2$ Ising Model}
\label{sec:rw}
Histogram-reweighting technique~\cite{Ferrenberg1988PRL,Ferrenberg1989PRL,Ferrenberg1995PRE} enables one to obtain continuous thermodynamic functions over a range of temperatures using data from only a single simulation. For the Hamiltonian in Eq.~(\ref{eq:Hamiltonian}), the total energy is
\begin{equation}
E = J_1\sum_{\langle i,j\rangle}\sigma_i\sigma_j + J_2\sum_{\langle\langle i,j\rangle\rangle}\sigma_i\sigma_j.
\end{equation}
The canonical partition function at inverse temperature $\beta=1/(k_B T)$ reads
\begin{equation}\label{eq:Z}
Z(\beta)=\sum_E N(E)\,e^{-\beta E},
\end{equation}
where $N(E)$ is the density of states.

We first choose a reference temperature $T_0$ with inverse temperature $\beta_0=1/(k_B T_0)$. Equilibrium configurations are sampled at $T_0$, and we build a two-dimensional histogram $H(E,m)$, where $E$ is the total energy and $m$ is the stripe magnetization defined in Eq.~\eqref{eq:magnetization_xy}. The normalized probability distribution under the reference temperature is
\begin{equation}
P_0(E,m)=\frac{H(E,m)}{\sum_{E,m} H(E,m)}.
\end{equation}

For a target temperature $T$ with $\beta=1/(k_B T)$, the reweighting rule gives
\begin{equation}\label{eq:reweight_prob}
P(E,m)=\frac{P_0(E,m)\,e^{-(\beta-\beta_0)E}}{\sum_{E,m} P_0(E,m)\,e^{-(\beta-\beta_0)E}}.
\end{equation}
For any thermodynamic observable $\mathcal O(E,m)$, its ensemble average under the target temperature is
\begin{equation}\label{eq:reweight_avg}
\langle\mathcal O\rangle_\beta =
\frac{\sum_{E,m} \mathcal O(E,m)\,H(E,m)\,e^{-(\beta-\beta_0)E}}
{\sum_{E,m} H(E,m)\,e^{-(\beta-\beta_0)E}}.
\end{equation}

Taking $\mathcal O=E$ gives the mean energy $\langle E\rangle_\beta$, and the specific heat follows from the energy fluctuations:
\begin{equation}
C_V = \frac{\langle E^2\rangle_\beta - \langle E\rangle_\beta^2}{N T^2}.
\end{equation}
For these energy-based observables, the magnetization can be marginalized out, leaving a one-dimensional energy histogram $H(E)=\sum_m H(E,m)$. Taking $\mathcal O=m^n$ gives the $n$-th moment of the magnetization. The Binder cumulant and susceptibility are then
\begin{equation}
Q = \frac{\langle m^4\rangle_\beta}{\langle m^2\rangle_\beta^2},
\qquad
\chi = \frac{N}{T}\left(\langle m^2\rangle_\beta - \langle m\rangle_\beta^2\right).
\end{equation}
These magnetization-based observables require the full two-dimensional histogram $H(E,m)$, since the mean values of $m^2$ and $m^4$ in each energy bin must be retained.

If the target temperature deviates too far from $T_0$, the exponential weights become concentrated on very few sampled configurations, inducing large statistical noise. Reliable results are therefore only obtainable within a finite temperature window around $T_0$. The main advantage of this approach is that a single simulation at $T_0$ yields thermodynamic curves over a continuous range of temperatures, eliminating the need to run separate simulations at each temperature of interest and substantially reducing the overall computational cost.

\section{Results}
\label{sec:res}
\subsection{Test Results of Algorithm Efficiency}
 
\begin{figure}[htbp]
\centering
\includegraphics[width=\linewidth]{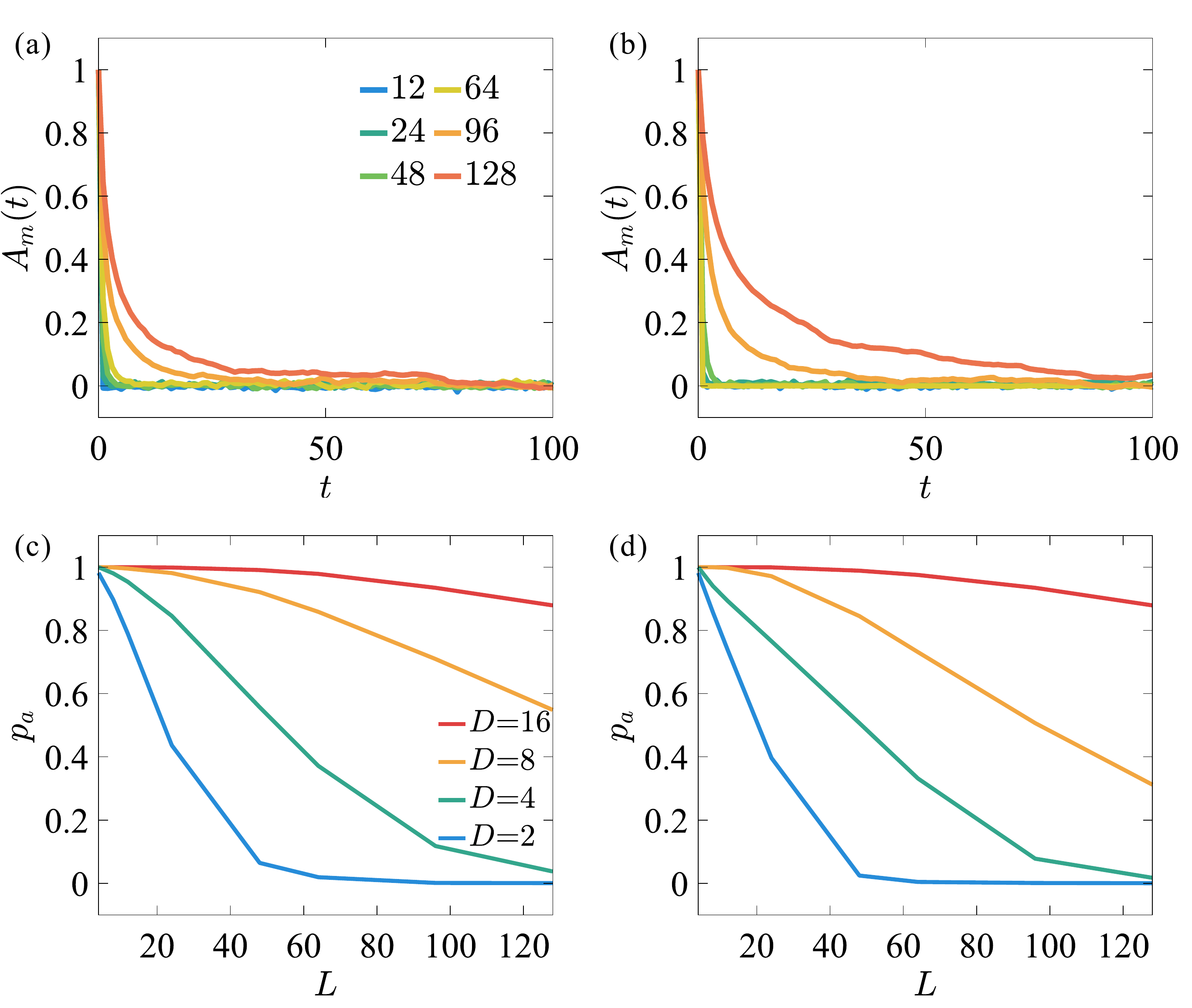}
\caption{Correlation time and acceptance probability of the TNMC method under open boundary conditions.
(a) Correlation time $A_m(t)$ at coupling strength $g=0.55$;
(b) Correlation time  $A_m(t)$ at coupling strength $g=1.0$;
(c) Acceptance probability \(p_{\text{a}}\) under different system sizes $L$ and bond dimensions $D$ for $g=0.55$ and $T=0.7713$;
(d) Acceptance probability \(p_{\text{a}}\) under different system sizes $L$ and bond dimensions $D$ for $g=1.0$ and $T$=2.083.}
\label{fig:ziyou}

\end{figure}

\begin{figure}[htbp]
\centering
\includegraphics[width=\linewidth]{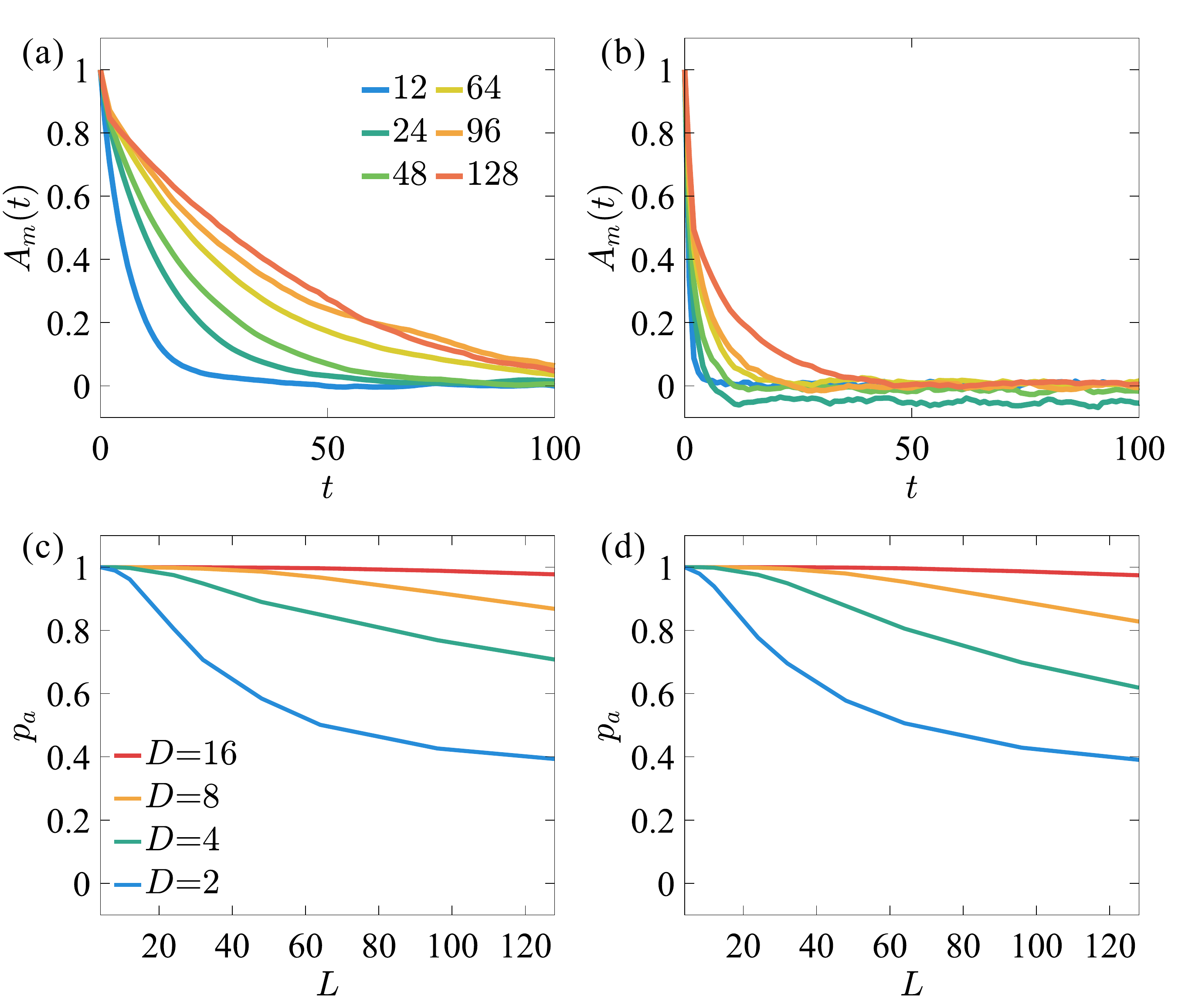}
\caption{Correlation time and acceptance probability of the TNMC method under periodic boundary conditions.
(a) Correlation time $A_m(t)$ obtained via TNMC for $g=0.55$;
(b) Correlation time $A_m(t)$ obtained via TNMC for $g=1.0$;
(c) Acceptance probability \(p_{\text{a}}\) under various system sizes $L$ and truncation dimensions $D$ at $g=0.55$;
(d) Acceptance probability \(p_{\text{a}}\) under various system sizes $L$ and truncation dimensions $D$ at $g=1.0$ and $T$=2.083.}
\label{fig:zhouqi}
\end{figure}

Under open boundary conditions, Fig~\ref{fig:ziyou} shows the correlation time $A_m(t)$ and acceptance probability \(p_{\text{a}}\) simulated by the TNMC method . Figures~\ref{fig:ziyou}  (a) and (b) plot the correlation function \(A(t)\) versus time $t$ at coupling strengths \(g=0.55\) and \(g=1.0\), respectively. Different curves represent different system sizes \(L = 12, 24, 48, 64, 96, 128\). The correlation function gradually decays with time, and its decay speed is affected by both the system size $L$ and the bond dimension $D$. By comparison, we can see that the correlation behavior changes when $g$ increases to \(1.0\).  Figures~\ref{fig:ziyou} (c) and (d) show how the acceptance probability \(p_{\text{a}}\) changes with system size $L$ for \(g=0.55\) and \(g=1.0\). For a fixed truncation dimension $D$ (from 2 to 16), the acceptance probability \(p_{\text{a}}\) decreases as the system size $L$ grows. In addition, a larger $D$ usually gives a higher acceptance probability, which shows the balance between calculation cost and sampling efficiency.

Under periodic  boundary conditions, Fig~\ref{fig:zhouqi} (a) and (b)  shows  $A_m(t)$     under different system sizes $L$ and truncation dimensions $D$ at \(g=0.55\) and \(g=1.0\) respectively.
The slower decay of the autocorrelation function under the block sweeping scheme, compared with open boundaries, is an inherent cost of the scheme rather than a sampling defect. First, the fixed boundary spins are inherited from the previous configuration and are not redrawn; near the weak first-order (or nearly continuous) transition at \(g=0.55\), they pin the large correlated domains, so the interior block update cannot fully randomize the domain orientation in a single sweep. Second, two alternating block updates [Figs.~\ref{fig:zhendang}(a) and~\ref{fig:zhendang}(f)] are required to refresh every spin, so each spin is updated only once every two sweeps, mechanically increasing the correlation time by roughly a factor of two when counted per block update. Finally, the four sites excluded from block updates are relaxed only by local Metropolis sweeps, which suffer from critical slowing down and feed their slower relaxation back into the network.

Fig~\ref{fig:zhouqi} (c) and (d)  show  \(p_{\text{a}}\) under periodic  boundary conditions.
Compared with open boundaries, the block sweeping scheme maintains a high acceptance probability for two reasons. Each block update generates a globally correlated configuration through the conditional probabilities \(P(\sigma_i|\sigma_{<i})\), which already account for the correlated environment and avoid crossing large energy barriers, while the fixed boundary spins further reduce the effective degrees of freedom and make the proposal distribution smoother and more concentrated.

\begin{figure}[t]
\centering
\includegraphics[width=\linewidth]{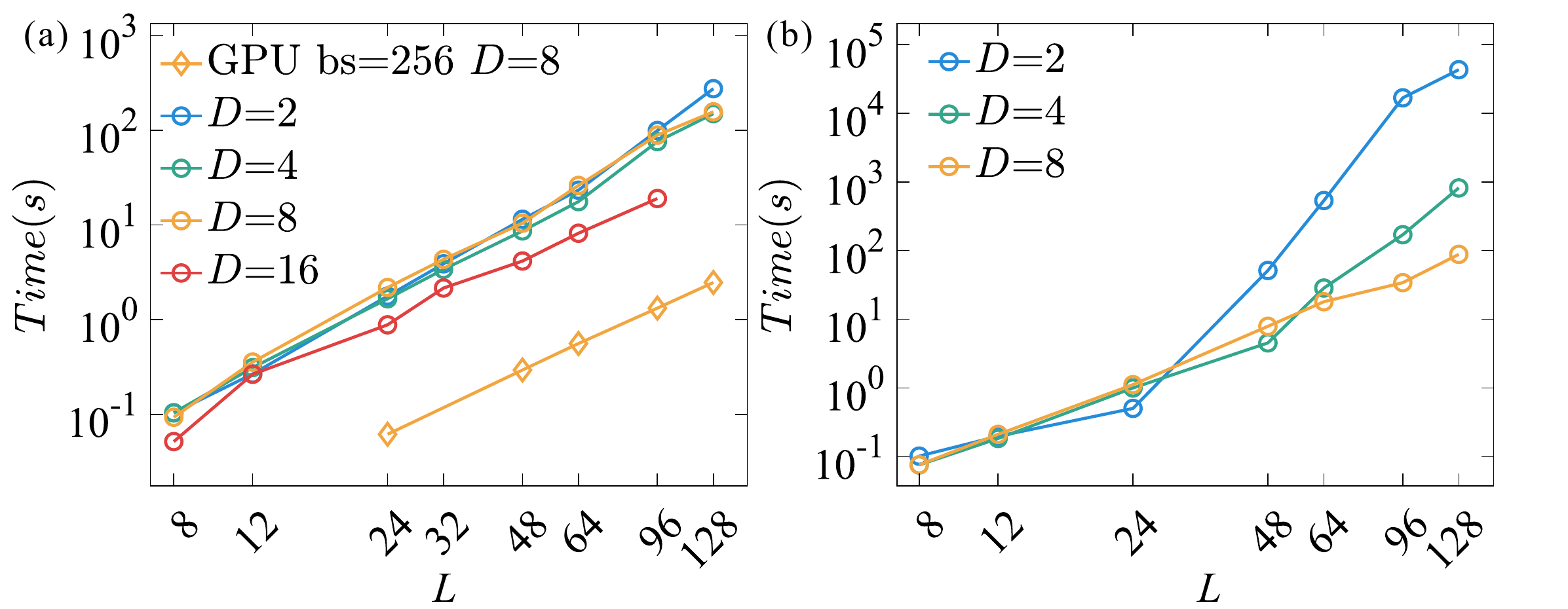} 
\caption{(a) Time cost for sampling one valid configuration versus system size $L$ under periodic boundary conditions with various truncation dimensions $D$, for both the CPU baseline and the GPU-accelerated results: the filled markers denote CPU results, while the orange diamond markers denote GPU results obtained on an NVIDIA GeForce RTX 3060 (12\,GB VRAM; $bs$ denotes batch size). GPU-accelerated sampling shortens the runtime by nearly two orders of magnitude compared to the CPU baseline.
(b) Time cost for sampling one valid configuration versus system size $L$ under open boundary conditions with truncation dimensions $D=2,4,8$.}
\label{fig:time}
\end{figure}
%Figure \ref{fig:time} plots the computation time of our TNMC algorithm against system size L. All axes are plotted on logarithmic scales.Panel (a) shows results under periodic boundary conditions with a fixed truncation dimension \(D=16\). The running time for sampling one valid configuration rises steadily as L grows from 8 to 96.Panel (b) presents data under open boundary conditions for three different truncation dimensions \(D=2,4,8\). At small lattice sizes (\(L \le 24\)), the three curves overlap closely with nearly identical time costs. As L increases further, obvious gaps appear between curves. Smaller D brings much higher time consumption, while a larger truncation dimension \(D=8\) keeps the sampling time far lower at the same system size.Overall, both boundary types lead to growing computation time with larger L. Under open boundaries, choosing a higher truncation dimension effectively reduces the total sampling overhead for big lattices.

Figure ~\ref{fig:time} presents the computational time cost for sampling a single valid spin configuration as a function of lattice size $L$, for both periodic and open boundary conditions with different tensor truncation dimensions $D$.
Both axes in Fig.~\ref{fig:time} are logarithmic, so an approximately straight line indicates polynomial scaling of the sampling time defined in Eq.~\ref{eq:time}, \(t\sim L^\lambda\), with the slope giving the exponent \(\lambda\). Under periodic block sweeping [panel (a)], the curves for different truncation dimensions remain close together: as $L$ increases from 8 to 128, the time per valid configuration grows only from about 10 s to the \(10^2\)--\(10^3\) s range, with a moderate slope of roughly \(1.5\)--2. The cost therefore follows a mild polynomial growth and remains tractable up to \(L=128\); the \(D=16\) curve is actually the least expensive at large $L$, because a larger $D$ yields a more accurate proposal and fewer wasted attempts. In contrast, the open-boundary curves [panel (b)] fan out markedly: at \(D=2\) the time reaches \(10^4\)--\(10^5\) s at \(L=128\) with a steeper slope of about \(2.5\)--3, and increasing $D$ lowers the cost. This follows because the time per valid configuration equals the time per sweep divided by the acceptance probability; at small $D$ under open boundaries the acceptance is low, so many sweeps are required to obtain one valid configuration, and the cost is dominated, and sharply amplified by this low acceptance. 

Recently, a GPU-based TNMC implementation, BatchTNMC~\cite{chen2025batchtnmcefficientsamplingtwodimensional}, has been introduced, which greatly accelerates sampling by parallelizing over multiple disorder realizations; for example, generating \(10^6\) samples at \(N=1024\times1024\) takes only a few hours on a single GPU. Since our system is a clean (non-disordered) model, we adapt the same batch-parallel strategy over independent samples; our code has been ported to GPU and its performance is tested.
Within batch size 256, running 500{,}000 sample cells takes roughly
$277{,}000$ seconds, or $3.2$ days, at $L=64$;
about $643{,}000$ seconds, or $7.4$ days, at $L=96$;
and approximately $1{,}182{,}000$ seconds, or $13.7$ days, at $L=128$.
Tests were conducted on a single NVIDIA GeForce RTX 3060
with $12\,\mathrm{GB}$ of VRAM, and upgrading the memory
and increasing the batch size would reduce the runtime nearly linearly,
bringing the above estimates down to a considerably shorter range.

\subsection{Results of Thermodynamic Observables}

\begin{figure}[t]
\centering
\includegraphics[width=\linewidth]{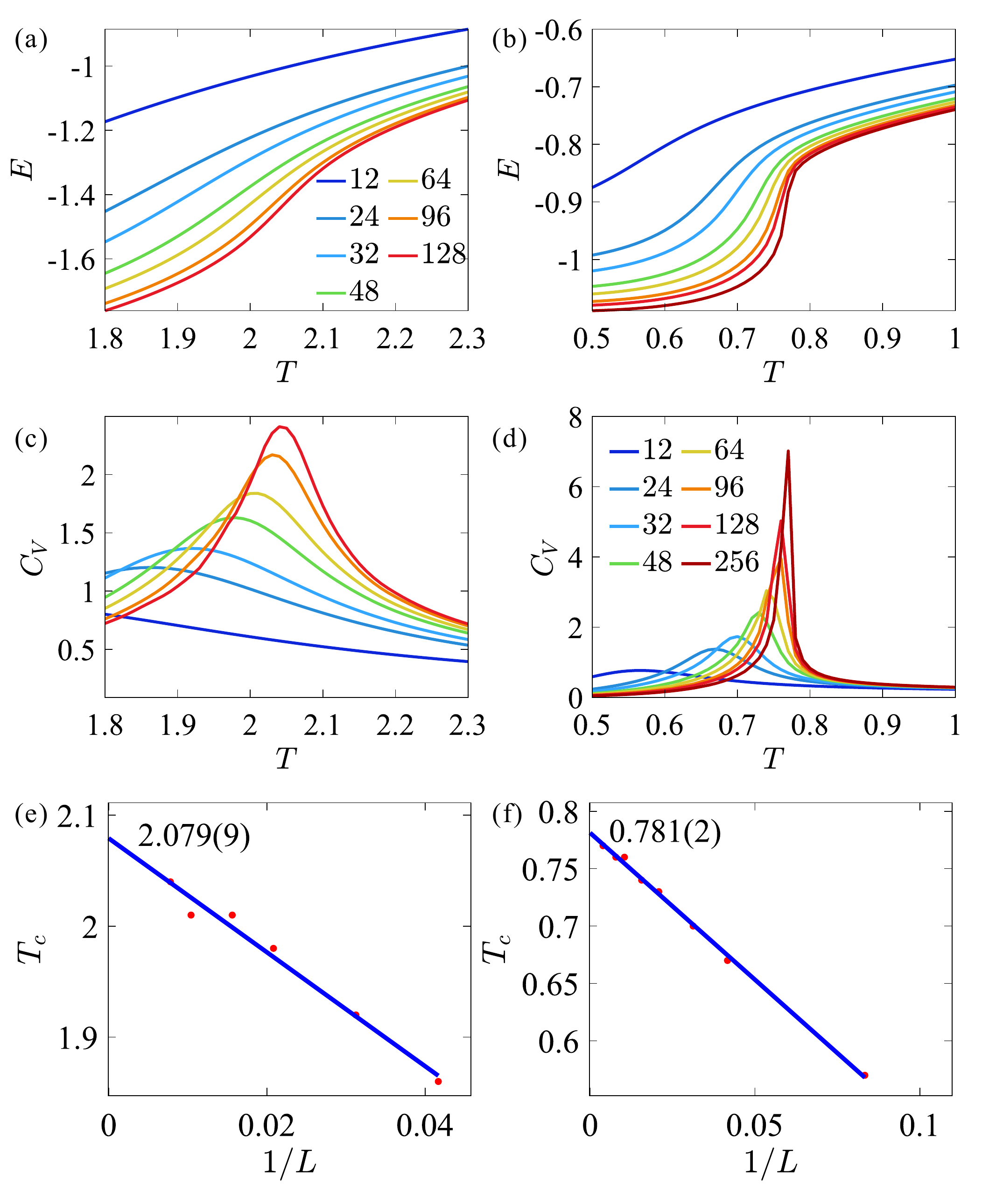}

\caption{Thermodynamic results and transition points directly calculated from the partition function for open boundary condition.
(a)(b) Curves of internal energy versus temperature for systems with different lattice sizes at coupling strengths $g=1.0$ and $g=0.55$, respectively;
(c)(d) Temperature dependence of specific heat $C_V$ for various lattice sizes at the two coupling strengths;
(e)(f) Linear finite-size extrapolations using the peak temperatures of specific heat for $g=1.0$ and $g=0.55$. In the thermodynamic limit $1/L \to 0$ ($L\to\infty$), the corresponding critical temperatures are $T_c\approx 2.079(9)$ and $T_c\approx 0.781(2)$.}

\label{fig:jieguo1}
\end{figure}

\subsubsection{Results from $\ln Z$}

For the \(J_1\)-\(J_2\) Ising model with open boundary conditions, we directly calculate physical quantities including energy and specific heat using the obtained partition function Z.
The corresponding formulas are given by
\begin{equation}
E = -\frac{\partial \ln Z}{\partial \beta},\quad C_V = \beta^2 \frac{\partial^2 \ln Z}{\partial \beta^2}
\label{eq:echCV}
\end{equation}
Only the first- and second-order numerical derivatives with respect to \(\beta\) are required.
This approach is computationally efficient. In particular, we compare the critical temperatures obtained here with those from the HOTRG method of previous work~\cite{hotrgj1j2023} at the two coupling strengths of interest, \(g=0.55\) and \(g=1.0\).

Figure \ref{fig:jieguo1} shows thermodynamic properties and critical behavior at \(g=1.0\) and \(g=0.55\). Panels (a)--(b) plot internal energy $E$ versus temperature for various lattice sizes $L$. The transition-region slope of $E$ steepens with increasing $L$, signalling a sharper phase transition.
Panels (c)--(d) display specific heat \(C_V\). Pronounced \(C_V\) peaks grow in height and shift with larger L, a hallmark of strong critical thermal fluctuations near the transition.
Panels (e)--(f) present finite-size scaling to extract the thermodynamic-limit critical temperature \(T_c\). We linearly extrapolate the specific-heat pseudo-critical temperatures against \(1/L\):
\begin{equation}
T_c(L) = T_c^\infty + \frac{A}{L}
\end{equation}
where \(T_c^\infty\) is the bulk critical temperature and $A$ is a constant correction amplitude.
The extrapolated results are \(T_c^\infty=0.781(2)\) (\(g=0.55\)) and \(T_c^\infty=2.079(9)\) (\(g=1.0\)).

For reference, large-scale HOTRG results~\cite{hotrgj1j2023} (\(L=32768\)) yield \(T_c=0.7713\) and \(T_c=2.083\). Notably, our sampling-free direct-\(\ln Z\) calculations, limited to \(L\le128\), produce extrapolated transition temperatures close to these large-lattice HOTRG benchmarks.
Small deviations remain, which may originate from our free-boundary setup, finite-size effects, and tensor-truncation errors. Further insight may be gained from forthcoming TNMC simulations, which can eliminate tensor-truncation errors.

\subsubsection{Comparison between histogram-reweighting and TNMC Results}

\begin{figure}[t]
\centering
\includegraphics[width=\linewidth]{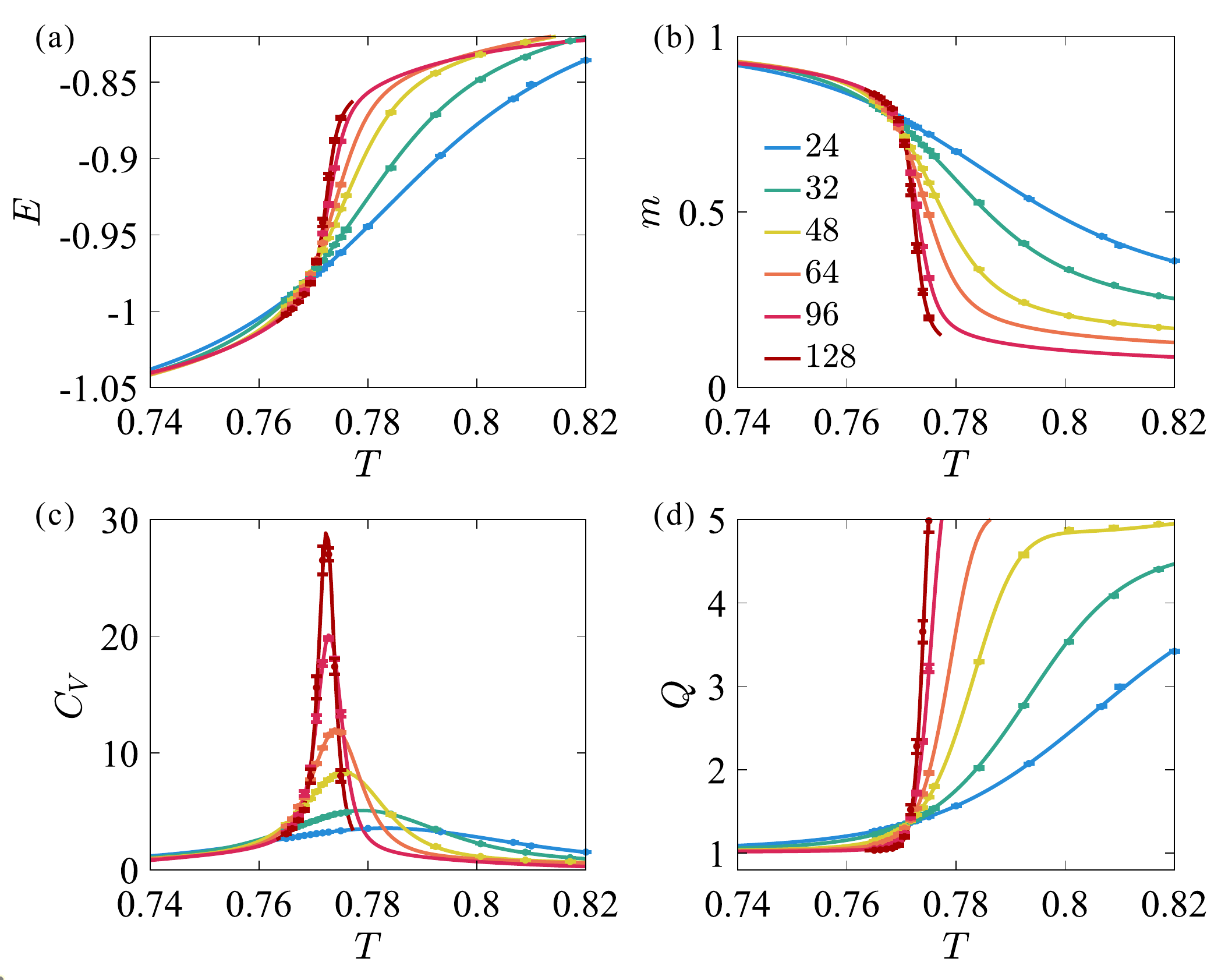}
\caption{Comparison of (a) the energy $E$, (b) the magnetization $m$, (c) the specific heat $C_V$, and (d) the Binder cumulant $Q$ obtained with the histogram-reweighting method (solid lines) and the TNMC method (data with error bars) under periodic boundary conditions at $g=0.55$, for system sizes $L=24,32,48,64,96,128$.}
\label{fig:TNMCMetro}
\end{figure}

\begin{figure}[t]
\centering
\includegraphics[width=\linewidth]{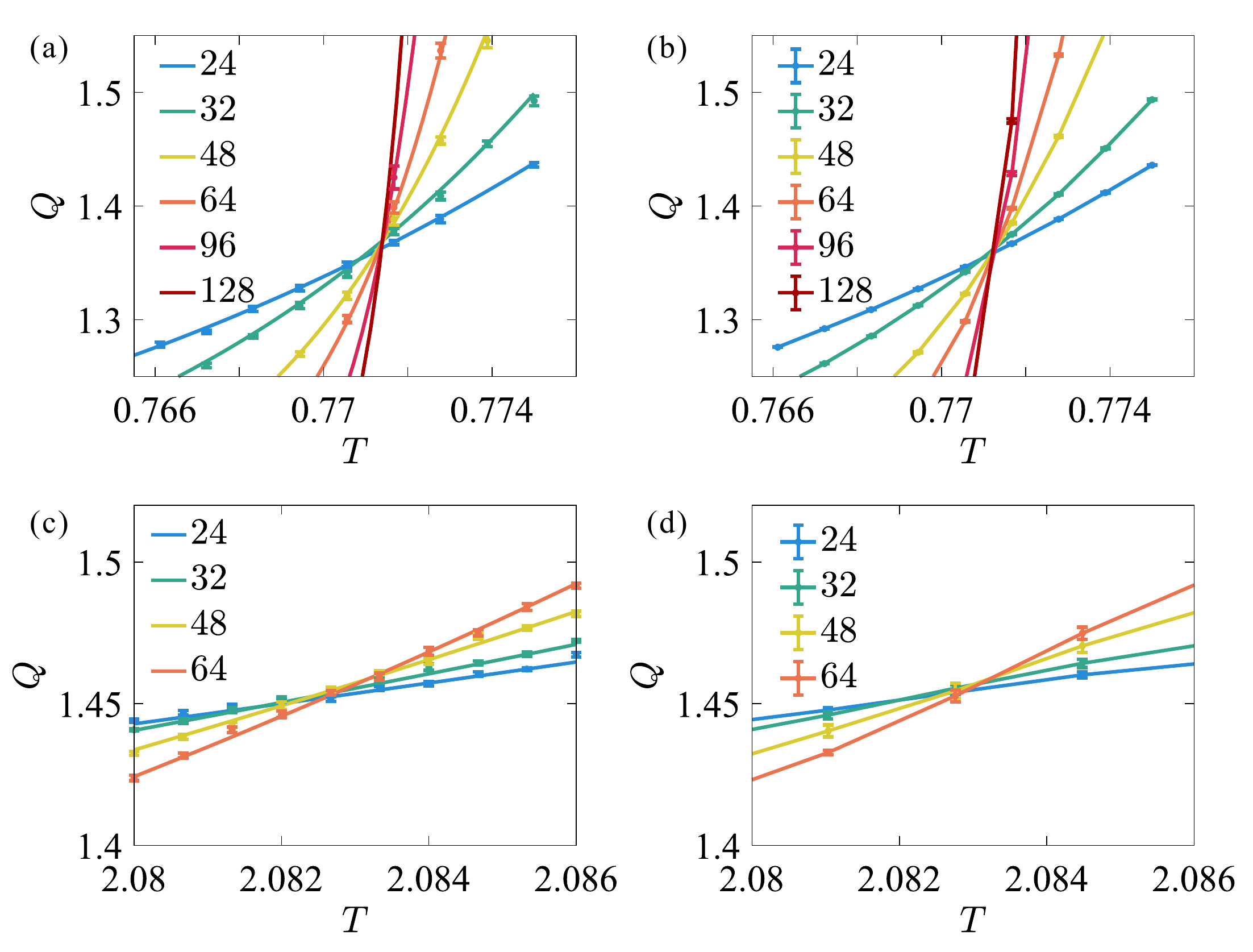}
\caption{Comparison of the Binder cumulant under periodic boundary conditions: (a) and (c) are obtained with the TNMC method (circles with error bars) and the histogram-reweighting method (solid lines), while (b) and (d) are obtained with the Metropolis method; (a) and (b) at $g=0.55$ for system sizes $L=24,32,48,64,96,128$, and (c) and (d) at $g=1.0$ for $L=24,32,48,64$.}
\label{fig:binder}
\end{figure}
Figure \ref{fig:TNMCMetro} compares thermodynamic quantities obtained with the TNMC and histogram reweighting~\cite{Ferrenberg1988PRL,Ferrenberg1989PRL} under periodic boundary conditions, at coupling strength \(g=0.55\) and for lattice sizes \(L=24,32,48,64,96,128\). Solid lines denote the histogram reweight results, whereas the discrete symbol markers with error bars are the TNMC data. Panels (a)--(d) correspond to the internal energy \(E\), the stripe magnetization \(m=\sqrt{m_x^2+m_y^2}\), the specific heat \(C_V\), and the Binder ratio \(Q\), respectively. We have also carried out full calculations using the standard Metropolis Monte-Carlo method; for conciseness these results are not plotted in the figure, yet they show consistent behavior with both TNMC and histogram-reweighting data within the considered parameter window. The TNMC data are in good agreement with the histogram reweighting results in the parameter window.

\subsubsection{More Precise Critical Points and Exponents}
% For more precise critical exponents, compute the Binder ratio for different sizes over a narrower temperature range.

To obtain more precise critical exponents, the Binder ratio needs to be computed for different system sizes over a narrower temperature range. Figure \ref{fig:binder} shows Binder cumulant \(Q(T)\) from TNMC and histogram-reweighting; in panels (a) and (c), the reweighted curves and the TNMC data points are in visually good agreement. The histogram-reweighting procedure is described in Appendix~\ref{app:reweight_impl}. Panels (a,b) correspond to \(g=0.55\) with lattice sizes \(L=24,32,48,64,96,128\), while panels (c,d) are for \(g=1.0\) with lattice sizes \(L=24,32,48,64\), where the largest system size is restricted to \(L=64\). For \(g=0.55\) , curves for distinct lattice sizes cross within the critical interval \(T\in[0.766,0.776]\). For \(g=1.0\), crossings are observed over \(T\in[2.08,2.086]\).

%To obtain more precise critical exponents, the Binder ratio needs to be computed for different system sizes over a narrower temperature range.
%Figure \ref{fig:binder}(a) shows the Binder cumulant \(Q\) as a function of temperature \(T\), computed with the TNMC algorithm at \(g=0.55\) under periodic boundary conditions. Curves are displayed for seven lattice sizes ranging from \(L=24\) to \(L=128\). With increasing temperature, the curves belonging to different sizes cross within the critical interval \(T\in[0.766,0.776]\).% According to finite-size scaling theory, the crossing point of the Binder cumulant locates the critical temperature of the thermodynamic limit, placing the transition of this system at \(T_c\approx0.771\).

\begin{table}[t]
\centering
\setlength{\tabcolsep}{2pt}
\caption{Summary of finite-size scaling fits. This table presents the selected core results; the complete set of fits, including different methods, various choices of $L_{\min}$, and fits both with and without the correction-to-scaling term, is provided in Appendix~\ref{app:binder}. At $g=0.55$, the fitted $y_t$ is consistent with weak first-order behavior~\cite{jin2012ashkin,jin2013prb}. For $g=1.0$, the Coulomb-gas prediction gives $y_t=1.20$~\cite{jin2012ashkin,jin2013prb}, which is reproduced by the Metropolis and TNMC fits within about 2 standard deviations.}
\begin{tabular}{lccccc}
\hline
\hline
$g$ & $L_{\min}/L_{\max}$ & $y_t=1/\nu$ & $T_c$ & $\chi^2/\mathrm{dof}$ & Method\\
\hline
0.55 & 24/128 & $1.686(8)$ & $0.771417(9)$ & 4.19 & Metropolis\\
0.55 & 32/128 & $1.687(7)$ & $0.771393(3)$ & 1.8 & Metropolis\\
0.55 & 24/96 & $1.69(3)$ & $0.77143(5)$ & 0.63 & TNMC\\
0.55 & 32/96 & $1.72(5)$ & $0.77143(7)$ & 0.61 & TNMC\\
% 0.55 & 24/96 & $1.67(3)$ & $0.77136(1)$ & 0.77 & Reweighting\\
% 0.55 & 32/96 & $1.67(4)$ & $0.7714(1)$ & 0.02 & Reweighting\\
1.0  & 24/128 & $1.24(3)$ & $2.083(2)$ & 0.75 & Metropolis\\
1.0  & 32/128 & $1.23(4)$ & $2.082(3)$ & 0.57 & Metropolis\\
1.0  & 24/96 & $1.17(2)$ & $2.08401(7)$ & 1.69 & TNMC\\
1.0  & 32/96 & $1.16(2)$ & $2.08422(9)$ & 1.35 & TNMC\\
% 1.0  & 24/128 & $1.17(1)$ & $2.083(3)$ & 4.93 & Reweighting\\
% 1.0  & 32/128 & $1.17(2)$ & $2.083(3)$ & 2.48 & Reweighting\\
\hline
\hline
\end{tabular}
\label{tab:Lmin_fit_two}
\end{table}

Using the finite size scaling equation,
\begin{equation}
Q = Q_0 + a_1(T - T_c)L^{y_t} + a_2(T - T_c)^2 L^{2y_t} + f_1 L^{-\omega},
\label{eq:q}
\end{equation}
 where, $Q_0$ is the critical Binder cumulant, $a_1,a_2$ are nonuniversal thermal-scaling amplitudes, $f_1$ is the correction-to-scaling amplitude, and $L$ is the lattice size. For fits with a relatively large $\chi^2/\mathrm{dof}$, we additionally include the third-order thermal-scaling term $a_3(T-T_c)^3L^{3y_t}$.

Table \ref{tab:Lmin_fit_two} summarizes the finite-size scaling fits obtained for different values of \(L_{\min}\). Fits use \(L\ge L_{\min}\in\{24,32\}\). Where the fits differ, we quote the result whose \(\chi^2/\mathrm{dof}\) is closest to unity. The TNMC results for \(g=0.55\) give \(y_t = 1.69(3)\), and the Metropolis results give \(y_t = 1.687(7)\); both lie somewhat below the standard first-order value \(y_t=2\). This suggests that, should the transition at \(g=0.55\) be weak first order, considerably larger system sizes would be needed for the exponent to converge to the standard first-order scaling behavior \(y_t=2\). The fitted critical points, \(T_c=0.77143(5)\) (TNMC) and \(T_c=0.771393(3)\) (Metropolis), are also very close to the value \(T_c=0.7713\) obtained with the HOTRG method at system size \(L=32768\)~\cite{hotrgj1j2023}. For \(g=1.0\), the TNMC and Metropolis fits give \(y_t=1.16(2)\) and \(y_t=1.24(3)\), respectively, consistent with the Coulomb-gas prediction \(y_t=1.20\)~\cite{jin2012ashkin,jin2013prb} within about 2 standard deviations; the corresponding critical points, \(T_c=2.08422(9)\) (TNMC) and \(T_c=2.083(2)\) (Metropolis), are close to the HOTRG value \(T_c=2.083\)~\cite{hotrgj1j2023}, and are also consistent with the transfer-matrix result \(T_c=2.08208(22)\) obtained from \(1/T_c=0.48029(5)\)~\cite{jin2013prb}. A detailed discussion of the fit results, including the dependence on \(L_{\min}\), the correction-to-scaling term, and the comparison among the two methods, is provided in Appendix~\ref{app:binder}.

\FloatBarrier
\section{Conclusion and Discussion}
\label{sec:con}
In summary, we have extended the TNMC algorithm to the two-dimensional frustrated $J_1$-$J_2$ Ising model with both open and periodic boundary conditions, and developed a block-sweeping scheme for periodic lattices. To save computational resources, we have also successfully coupled the sampler with histogram reweighting to determine the phase transition point.
Our numerical simulations demonstrate that, compared with the conventional Metropolis algorithm, TNMC essentially removes critical slowing down, reduces the correlation between samples, and improves the sampling efficiency. 

The finite-size scaling analysis yields critical temperatures consistent with the HOTRG~\cite{hotrgj1j2023} and Metropolis~\cite{jin2012ashkin,jin2013prb} results at both $g=0.55$ and $g=1.0$.
These findings are relevant to the ongoing debate on the nature of the phase transitions in this model~\cite{frustrated_book,Moran-Lopez1993,AguileraGranja1993,Moran-Lopez1994,Lopez-Sandoval1999,dosAnjos2008,Kalz2011,jin2012ashkin,jin2013prb,li2021tensor,sun2026generalized,Binder1980,landau1980,Oitmaa1981,gangat2024,lijiang2024ptep,chatelain2024,kalz2012potts,hucharbonneau2021,abalmasov2023,leekimkim2024}: at $g=1.0$, the fitted exponent is close to the Coulomb-gas value, supporting a continuous transition, while at $g=0.55$ the exponents below the first-order value are consistent with weak first-order behavior, which would require considerably larger system sizes to be confirmed~\cite{lijiang2024ptep}. Given the accessible system sizes, we refrain from a definitive statement on the transition order.

The TNMC algorithm established here provides a promising tool for future investigations of this frustrated system. Simulations at such large sizes are computationally demanding, but GPU acceleration and batch sampling~\cite{chen2025batchtnmcefficientsamplingtwodimensional} could substantially reduce the actual running time.

% The advantages and limitation of the present approach are as follows.
% Correlations between samples generated by this method are weak, which effectively mitigates the critical slowing down inherent to the Metropolis algorithm. The block-sweeping scheme enables TNMC to efficiently simulate systems with periodic boundary conditions and suppresses deviations of critical exponents induced by boundary effects. Although the runtime is higher than that of the Metropolis method, this drawback can be addressed using state-of-the-art GPU acceleration~\cite{chen2025batchtnmcefficientsamplingtwodimensional}.

{\it Acknowledgements.} W.Z. is grateful to Dr. E. D. Guo for simulation discussion. W. Z. was supported by
the Hefei National Research Center for Physical Sciences
at the Microscale (Grant No. KF2021002) and the Fundamental Research Program of Shanxi Province (Grant
Nos. 202303021221029).
Y.D. was supported by the National Natural
Science Foundation of China under Grant No. 12275263
and the Quantum Science and Technology National
Science and Technology Major Project under Grant
No. 2021ZD0301900.
\appendix

\section{TNMC Markov Chain Algorithm}
\label{app:tnmc_algo}

For completeness, we summarize the TNMC Markov chain used in this work as follows.

\smallskip
\noindent\hangindent=2em 1. Build the tensor network on the rotated lattice with interaction tensors $T$ and copy tensors $\delta$ (Sec.~\ref{sec:method}).\par
\noindent\hangindent=2em 2. Precompute the upper tensors $A'_{y,x}$ by contracting rows from top to bottom.\par
\noindent\hangindent=2em 3. Precompute the right tensors $A''_{y,x}$ by contracting columns from right to left.\par
\noindent\hangindent=2em 4. Set $t=0$; initialize the spin configuration (ground state for warm-up, otherwise the previous configuration).\par
\noindent\hangindent=2em 5. If $t>t_{\max}$ go to step 11.\par
\noindent\hangindent=2em 6. Fix the boundary-width=2 ring of spins; sample the interior spins row by row from top to bottom, left to right.\par
\noindent\hangindent=2.8em \hspace*{2em}(a) At each site $(y,x)$, contract the cached upper, left, right, and lower tensors to obtain $P(\sigma_i=\pm1|\sigma_{<i})$ via Eq.~\eqref{eq:cp}.\par
\noindent\hangindent=2.8em \hspace*{2em}(b) Draw $\sigma_i$ from this Bernoulli distribution and project it into the network through the projection vector $\nu$.\par
\noindent\hangindent=2.8em \hspace*{2em}(c) Update the left cache (contracted leftward) and the lower cache (contracted downward); move to the next site.\par
\noindent\hangindent=2.8em \hspace*{2em}(d) After finishing a row, recompute the right tensors $A''_{y,x}$ for the next row.\par
\noindent\hangindent=2em 7. Swap the roles of the fixed boundary ring and the sampled interior [Figs.~\ref{fig:zhendang}(a) and \ref{fig:zhendang}(f)].\par
\noindent\hangindent=2em 8. Perform a single local Metropolis sweep over all $N$ spins to relax the four sites not covered by the block update.\par
\noindent\hangindent=2em 9. If $t>t_{\rm th}$, measure the physical observables (energy, magnetization, etc.).\par
\noindent\hangindent=2em 10. $t\leftarrow t+1$; go to step 5.\par
\noindent\hangindent=2em 11. End.\par
\section{Practical Implementation of Histogram Reweighting for the Binder Cumulant}
\label{app:reweight_impl}

The theoretical framework for histogram reweighting has been introduced in Sec.~\ref{sec:rw}. Here we describe the practical implementation used to obtain the Binder cumulant data analyzed in the finite-size scaling fits. The TNMC results in Table~\ref{tab:binder} are obtained from direct TNMC simulations carried out at each target temperature independently, while the Metropolis results are obtained from standard local-update Monte Carlo simulations.

For each system size $L$ and independent seed, the TNMC simulation at the reference temperature $T_0$ records four time series: the total energy $E$, the per-spin magnetization $m$ defined in Eq.~\eqref{eq:magnetization_xy}, and the second and fourth moments of the magnetization $m^2, m^4$. For the two parameter points considered, at $J_2/|J_1|=0.55$ we use $T_0=0.77167$ and scan $T\in[0.765, 0.775]$ with 50 points, while at $J_2/|J_1|=1.0$ we use $T_0=2.08533$ and scan $T\in[2.080, 2.086]$ with 20 points. System sizes are $L\in\{24,32,48,64\}$ with $N=L^2/2$ spins, and ten independent seeds are run for each $L$. For each seed, we discard the first 16,000 sweeps for thermalization and collect 160,000 configurations for measurement, giving 1,600,000 configurations in total for each $L$.

The workflow is as follows. For each $L$, we run $n_s=10$ independent TNMC simulations at $T_0$, recording $E, m, m^2, m^4$. We then concatenate all seeds and build the joint histogram $H(E,m)$, recording the mean values of $m^2$ and $m^4$ in each bin, reweight to each target $T$ in the scan window using the weight $w=\exp[(1/T-1/T_0)E]$, and compute the Binder cumulant $Q(T,L)$. To estimate statistical errors, we repeat this procedure for each leave-one-seed-out jackknife subset and obtain $\sigma_Q(T,L)$. The final $(L, T, Q, \sigma_Q)$ data are exported for finite-size scaling fitting. The reliable reweighting window is set by the energy overlap at $T_0$ and $T$~\cite{Ferrenberg1988PRL,Ferrenberg1995PRE}: approximately $|\Delta T|\lesssim T/\sqrt{C_V}$, where $C_V$ is the total specific heat at $T_0$. For the present system sizes and the narrow windows used ($|\Delta T|\lesssim0.005$ at $J_2/|J_1|=1.0$, $|\Delta T|\lesssim0.007$ at $J_2/|J_1|=0.55$), which are well below the upper bound $|\Delta T|\lesssim T_0/\sqrt{C_V}\sim0.1$--$0.4$ estimated from the measured specific heat.

\section{Binder Cumulant Scaling Results at $J_2/|J_1|=0.55$ and $1.0$}
\label{app:binder}

\begin{table*}[!t]
\centering
\footnotesize
\setlength{\tabcolsep}{4pt}
\begin{tabular}{l@{\hspace{2em}}lccccccccc}
\toprule
Method & min $L$ & max $L$ & $Q_0$ & $T_c$ & $y_t$ & $a_1$ & $a_2$ & $f_1$ & $\omega$ & $\chi^2/\mathrm{dof}$ \\
\midrule
\multicolumn{11}{l}{$J_2/|J_1|=0.55$}\\
Metropolis & 24 & 128 & 1.361(1) & 0.7713(3) & 1.700(3) & 0.083(1) & 0.010(2) & --- & --- & 80.8 \\
Metropolis & 32 & 128 & 1.366(2) & 0.7714(3) & 1.704(4) & 0.083(1) & 0.009(3) & --- & --- & 18.8 \\
Metropolis & 48 & 128 & 1.370(4) & 0.7714(4) & 1.707(6) & 0.082(2) & 0.009(4) & --- & --- & 11.6 \\
% Reweighting & 24 & 96 & 1.3647(5) & 0.77136(1) & 1.67(3) & 0.092(9) & 0.014(3) & --- & --- & 0.77 \\
% Reweighting & 32 & 96 & 1.3696(9) & 0.7714(1) & 1.67(4) & 0.091(1) & 0.012(3) & --- & --- & 0.02 \\
% Reweighting & 48 & 96 & 1.369(2) & 0.7714(1) & 1.68(5) & 0.090(2) & 0.012(5) & --- & --- & 0.02 \\
TNMC       & 24 & 96 & 1.362(1) & 0.77128(3) & 1.63(3) & 0.10(1) & 0.014(3) & --- & --- & 1.2 \\
TNMC       & 32 & 96 & 1.365(2) & 0.77133(3) & 1.67(3) & 0.091(1) & 0.010(2) & --- & --- & 0.75 \\
Metropolis & 24 & 128 & 1.372(3) & 0.77136(2) & 1.703(9) & 0.083(3) & 0.0091(6) & -48(117) & 2.7(8) & 7.46 \\
Metropolis & 24 & 128 & 1.385(6) & 0.7714(4) & 1.714(3) & 0.080(9) & 0.008(2) & -0.53(1) & 1 & 11.1 \\
Metropolis & 32 & 128 & 1.38(1) & 0.7714(6) & 1.718(4) & 0.079(1) & 0.008(3) & -0.36(3) & 1 & 11.5 \\
Metropolis & 48 & 128 & 1.365(3) & 0.7714(1) & 1.69(1) & 0.086(4) & 0.010(8) & 0.2(1) & 1 & 12.9 \\
% Reweighting & 24 & 96 & 1.384(3) & 0.77146(2) & 1.69(3) & 0.085(9) & 0.009(2) & -0.47(7) & 1 & 0.16 \\
% Reweighting & 32 & 96 & 1.369(6) & 0.77140(3) & 1.67(4) & 0.092(1) & 0.013(4) & 0.03(16) & 1 & 0.02 \\
% Reweighting & 48 & 96 & 1.38(2) & 0.77143(6) & 1.69(6) & 0.084(2) & 0.010(5) & -0.5(7) & 1 & 0.009 \\
TNMC       & 24 & 96 & 1.383(6) & 0.77143(5) & 1.69(3) & 0.087(9) & 0.008(2) & -0.4(1) & 1 & 0.63 \\
TNMC       & 32 & 96 & 1.38(1) & 0.77143(7) & 1.72(5) & 0.07(1) & 0.006(2) & -0.5(3) & 1 & 0.61 \\
%TNMC       & 48 & 128 & 1.42(3) & 0.772(1) & 1.82(9) & 0.052(2) & 0.003(2) & -2.1(13) & 1 & 0.59 \\
\midrule
\multicolumn{11}{l}{$J_2/|J_1|=1.0$}\\
Metropolis & 24 & 128 & 1.454(1) & 2.083(1) & 1.25(3) & 0.066(1) & 0.005(2) & --- & --- & 0.71 \\
Metropolis & 32 & 128 & 1.455(1) & 2.083(1) & 1.23(4) & 0.070(1) & 0.007(3) & --- & --- & 0.79 \\
Metropolis & 48 & 128 & 1.452(2) & 2.083(2) & 1.22(6) & 0.074(2) & 0.006(4) & --- & --- & 0.87 \\
% Reweighting & 24 & 128 & 1.454(2) & 2.083(3) & 1.17(1) & 0.089(4) & 0.001(2) & --- & --- & 4.93 \\
% Reweighting & 32 & 128 & 1.458(3) & 2.083(3) & 1.17(2) & 0.087(6) & 0.006(2) & --- & --- & 2.48 \\
% Reweighting & 48 & 128 & 1.462(4) & 2.084(4) & 1.17(2) & 0.088(8) & 0.010(2) & --- & --- & 0.32 \\
% TNMC       & 24 & 64 & 1.4538(4) & 2.0826(8) & 1.03(3) & 0.12(1) & 0.03(2) & --- & --- & 15.4 \\
% TNMC       & 32 & 64 & 1.4554(5) & 2.08287(9) & 1.15(4) & 0.07(1) & 0.006(6) & --- & --- & 19.8 \\
TNMC       & 24 & 96 & 1.4565(2) & 2.08315(3) & 1.16(1) & 0.092(6) & 0.006(2) & --- & --- & 6.26 \\
TNMC       & 32 & 96 & 1.4591(3) & 2.08335(4) & 1.15(2) & 0.098(7) & 0.009(2) & --- & --- & 4.16 \\
% TNMC       & 24 & 128 & 1.4587(2) & 2.08348(3) & 1.22(1) & 0.069(4) & 0.0045(7) & --- & --- & 20.76 \\
% TNMC       & 32 & 128 & 1.4618(3) & 2.08372(3) & 1.22(2) & 0.070(5) & 0.0054(8) & --- & --- & 18.75 \\
TNMC       & 24 & 96 & 70(6000) & 2.0847(2) & 1.19(2) & 0.084(7) & 0.008(1) & -68(5987)  & 0.000(21) & 1.50 \\
Metropolis & 24 & 128 & 1.453(3) & 2.083(2) & 1.24(3) & 0.066(9) & 0.005(3) & 0.02(7) & 1 & 0.75 \\
Metropolis & 32 & 128 & 1.445(5) & 2.082(3) & 1.23(4) & 0.070(1) & 0.004(3) & 0.3(1) & 1 & 0.57 \\
% Reweighting & 24 & 128 & 1.471(9) & 2.084(5) & 1.18(1) & 0.086(5) & 0.009(2) & -0.36(2) & 1 & 0.23 \\
% Reweighting & 32 & 128 & 1.477(2) & 2.084(7) & 1.19(2) & 0.083(6) & 0.009(2) & -0.50(4) & 1 & 0.01 \\
% TNMC       & 24 & 64 & 1.461(3) & 2.0832(2) & 1.03(3) & 0.12(1) & 0.01(1) & -0.13(4) & 1 & 15.6 \\
% TNMC       & 32 & 64 & 1.446(5) & 2.0822(4) &1.13(4) & 0.08(1) & 0.012(8) & 0.2(1) & 1 & 20.5 \\
% TNMC       & 24 & 128 & 1.484(1) & 2.08461(6) & 1.21(2) & 0.076(5) & 0.0067(8) & -0.59(2) & 1 & 8.54 \\
% TNMC       & 32 & 128 & 1.496(2) & 2.08505(8) &1.21(2) & 0.077(6) & 0.0067(8) & -0.96(5) & 1 & 7.82 \\
TNMC       & 24 & 96 & 1.473(1) & 2.08401(7) & 1.17(2) & 0.090(6) & 0.010(1) & -0.37(3) & 1 & 1.69 \\
TNMC       & 32 & 96 & 1.479(2) & 2.08422(9) &1.16(2) & 0.094(8) & 0.011(2) & -0.51(5) & 1 & 1.35 \\
\bottomrule
\end{tabular}
\caption{Fitted Binder cumulant scaling parameters at $J_2/|J_1|=0.55$ and $1.0$, with and without the correction-to-scaling term $f_1 L^{-\omega}$.}
\label{tab:binder}
\end{table*}

\begin{table*}[!t]
\centering
\footnotesize
\setlength{\tabcolsep}{2pt}
\begin{tabular}{l@{\hspace{1em}}lcccccccccc}
\toprule
Method & min $L$ & max $L$ & $Q_0$ & $T_c$ & $y_t$ & $a_1$ & $a_2$ & $a_3$ & $f_1$ & $\omega$ & $\chi^2/\mathrm{dof}$ \\
\midrule
\multicolumn{11}{l}{$J_2/|J_1|=0.55$}\\
Metropolis & 24 & 128 & 1.3695(9) & 0.771382(6) & 1.6763(5) & 0.09043(14) & 0.01209(5) & 0.0007(1) & -92(12) & 2.940(44) & 1.43 \\
% Metropolis & 24 & 96 & 1.3698(12) & 0.771386(9) & 1.674(4) & 0.0912(11) & 0.0122(4) & 0.0007(1) & -76(100) & 2.87(46) & 0.82 \\
% TNMC       & 24 & 128 & 1.3698(12) & 0.771386(9) & 1.674(4) & 0.0912(11) & 0.0122(4) & 0.0007(1) & -76(104) & 2.87(46) & 0.82 \\
% TNMC       & 24 & 96 & --- & 0.77131(9) & 1.60(5) & 0.118(22) & 0.0206(99) & 0.0016(21) & --- & 0.000(4) & 1.29 \\
Metropolis & 24 & 128 & 1.380(2) & 0.771417(9) & 1.686(8) & 0.088(2) & 0.0112(8) & 0.0006(4) & -0.43(3) & 1 & 4.19 \\
Metropolis & 32 & 128 & 1.374(2) & 0.771393(7) & 1.687(7) & 0.087(2) & 0.0114(6) & 0.0007(13) & -0.25(4) & 1 & 1.80 \\
Metropolis & 24 & 96 & 1.383(1) & 0.771444(9) & 1.684(6) & 0.089(2) & 0.0113(6) & 0.0007(2) & -0.48(3) & 1 & 2.10 \\
Metropolis & 32 & 96 & 1.376(2) & 0.771404(9) & 1.684(6) & 0.088(1) & 0.0115(4) & 0.00065(9) & -0.28(4) & 1 & 0.79 \\
% TNMC       & 24 & 128 & 1.381(5) & 0.77143(4) & 1.67(3) & 0.093(1) & 0.011(3) & 0.0003(5) & -0.44(12) & 1 & 0.64 \\
% TNMC       & 32 & 128 & 1.379(9) & 0.77141(5) & 1.69(4) & 0.08(1) & 0.00985(4) & 0.0002(4) & -0.39(15) & 1 & 0.63 \\
TNMC       & 24 & 96 & 1.383(5) & 0.77144(4) & 1.67(3) & 0.09(1) & 0.011(3) & 0.0003(5) & -0.47(12) & 1 & 0.65 \\
TNMC       & 32 & 96 & 1.382(10) & 0.77143(6) & 1.71(5) & 0.08(2) & 0.008(4) & 0.0002(3) & -0.48(18) & 1 & 0.65 \\

\midrule
\multicolumn{11}{l}{$J_2/|J_1|=1.0$}\\
% Metropolis & 24 & 128 & --- & 2.0826(4) & 1.242(47) & 0.07(1) & 0.005(4) & -0.0001(2) & --- & --- & 0.82 \\
Metropolis & 24 & 128 & 1.453(3) & 2.0827(2) & 1.24(6) & 0.06(1) & 0.005(4) & -0.00001(24) & 0.02(6) & 1 & 0.79 \\
Metropolis & 32 & 128 & 1.445(4) & 2.0824(2) & 1.22(5) & 0.07(2) & 0.004(4) & 0.0007(37) & 0.25(10) & 1 & 0.62 \\
% TNMC       & 24 & 128 & 1.484(3) & 2.0847(2) & 1.22(5) & 0.08(2) & 0.021(10) & 0.008(6) & -0.59(7) & 1 & 7.88 \\
% TNMC       & 32 & 128 & 1.496(5) & 2.0851(2) & 1.23(5) & 0.08(2) & 0.018(9) & 0.005(4) & -0.95(12) & 1 & 7.07 \\
TNMC       & 24 & 96 & 1.473(2) & 2.0840(9) & 1.17(3) & 0.092(10) & 0.012(8) & 0.002(8) & -0.37(4) & 1 & 1.73 \\
TNMC       & 32 & 96 & 1.477(2) & 2.0842(1) & 1.16(3) & 0.096(11) & 0.013(8) & 0.002(8) & -0.52(6) & 1 & 1.39 \\

\bottomrule
\end{tabular}
\caption{Fitted Binder cumulant scaling parameters at $J_2/|J_1|=0.55$ and $1.0$, with and without the correction-to-scaling term $f_1 L^{-\omega}$. In contrast to Table~\ref{tab:binder}, which keeps only up to the second-order correction terms ($a_1$, $a_2$, $f_1$), the fits here additionally include the third-order correction term $a_3(T-T_c)^3 L^{3y_t}$ in Eq.~\ref{eq:q}.}
\label{tab:binder3}
\end{table*}

Several observations can be drawn from Table~\ref{tab:binder}. First, the critical coupling $T_c$ (in the units of $|J_1|/k_B$) is remarkably stable within each method: at $J_2/|J_1|=0.55$, once the correction-to-scaling term is included, it converges to $T_c=0.7714(4)$ for the Metropolis fits and $T_c=0.77143(5)$ for the TNMC fits, while at $J_2/|J_1|=1.0$ the corresponding values are $T_c=2.083(2)$ (Metropolis) and $T_c=2.08422(9)$ (TNMC). This convergence across the two independent simulation techniques (Metropolis and TNMC) provides a strong consistency check on the critical temperature.

Second, the thermal exponent $y_t$ depends more sensitively on the fit scheme. At $J_2/|J_1|=0.55$, the Metropolis data give $y_t=1.707(6)$ without the correction-to-scaling term and $y_t=1.714(3)$ with it ($\omega=1$). The TNMC results at this parameter point give $y_t=1.63(3)$ without the correction term and $y_t=1.69(3)$ with it ($\omega=1$).
Within their respective error bars, both methods are mutually consistent and yield $y_t\approx1.69$--$1.72$, somewhat below the standard first-order value $y_t=2$. As pointed out in Ref.~\cite{jin2012ashkin,jin2013prb}, this behavior is consistent with a weakly first-order transition, and the fitted exponent is expected to approach $y_t=2$ only on considerably larger lattices.

Third, at $J_2/|J_1|=1.0$, the Metropolis fits give $y_t=1.22(6)$, while the TNMC fits yield $y_t=1.16(2)$, both approximately consistent with the Coulomb-gas prediction $y_t=1.20$~\cite{jin2012ashkin,jin2013prb} within about 2 standard deviations.

Moreover, the $\chi^2/\mathrm{dof}$ values for some of the fits in Table~\ref{tab:binder} are larger than unity. At $J_2/|J_1|=0.55$, the Metropolis fits without corrections give $\chi^2/\mathrm{dof}$ as large as $80.8$, $18.8$, and $11.6$, which remain somewhat above unity ($11.1$--$12.9$) even with the $\omega=1$ term. At $J_2/|J_1|=1.0$, the Metropolis fits yield $\chi^2/\mathrm{dof}\approx0.6$--$0.9$, whereas the TNMC fits give $6.26$ and $4.16$ without corrections, which reduce to $1.69$ and $1.35$ with the $\omega=1$ term. This is possibly because only the leading correction terms such as $a_1$, $a_2$, and $f_1$ were included in the fit, which may be insufficient to account for the residual finite-size corrections.

To further examine the finite-size corrections and improve the quality of the fits, we extend the analysis to include higher-order correction terms; the results are collected in Table~\ref{tab:binder3}. For the Metropolis data at $J_2/|J_1|=0.55$, treating the exponent $\omega$ as a free parameter is not reliable, because the statistical errors of several fitted quantities exceed the values themselves; following common practice in finite-size scaling analyses of this model~\cite{chen2025tensor}, we fix $\omega=1$. With these higher-order terms, the $\chi^2/\mathrm{dof}$ values are all reasonable, remaining close to unity and ranging from $0.62$ to $4.19$. The fitted exponents and transition temperatures are essentially consistent with the corresponding values in Table~\ref{tab:binder}.

\bibliographystyle{apsrev4-2-withtitle}
\bibliography{j1j2}

\end{document}